\documentclass[aps,prX,superscriptaddress,twocolumn]{revtex4}
\usepackage{amsmath}
\usepackage{amsfonts}
\usepackage{enumerate}
\usepackage{graphicx}
\usepackage{color}
\usepackage{dsfont}
\usepackage{multirow}
\usepackage{verbatim}
\usepackage{mathtools}
\usepackage[normalem]{ulem}
\usepackage{comment}
\usepackage[colorlinks=true,citecolor=blue,linkcolor=magenta]{hyperref}

\begin{document}

\newcommand{\beq}{\begin{equation}}
\newcommand{\eeq}{\end{equation}}
\newcommand{\beqa}{\begin{eqnarray}}
\newcommand{\eeqa}{\end{eqnarray}}
\newcommand{\ben}{\begin{enumerate}}
\newcommand{\een}{\end{enumerate}}
\newcommand{\hs}{\hspace{1.5mm}}
\newcommand{\vs}{\vspace{0.5cm}}
\newcommand{\note}[1]{{\color{red}  #1}}
\newcommand{\notea}[1]{{\bf #1}}
\newcommand{\new}[1]{{#1}}
\newcommand{\ket}[1]{|#1 \rangle}
\newcommand{\bra}[1]{\langle #1|}
\newcommand{\im}{\dot{\imath}}
\newcommand{\tg}[1]{\textcolor{blue}{#1}}
\newcommand{\f}{f^{\phantom{\dagger}}}
\newcommand{\rB}{\boldsymbol{r}}
\newcommand{\kB}{\boldsymbol{k}}
\newcommand{\Tr}{\mathrm{Tr}}
\newcommand{\mS}{\mathcal{S}}
\newcommand{\mU}{\mathcal{U}}
\newcommand{\mO}{\mathcal{O}}

\title{Anomalous Subdiffusion from Subsystem Symmetries }

\author{Jason Iaconis}
\affiliation{Department of Physics and Center for Theory of Quantum Matter, University of Colorado, Boulder, Boulder, Colorado 80309, USA}
\author{Sagar Vijay}
\thanks{Corresponding Author: \underline{\href{mailto:sagarvijay@fas.harvard.edu}{sagarvijay@fas.harvard.edu}}}
\affiliation{Department of Physics, Harvard University, Cambridge, MA 02318, USA }
\author{Rahul Nandkishore}
\affiliation{Department of Physics and Center for Theory of Quantum Matter, University of Colorado, Boulder, Boulder, Colorado 80309, USA}


\begin{abstract}
We introduce quantum circuits in two and three spatial dimensions which are classically simulable, despite producing a high degree of operator entanglement. We provide a partial characterization of these ``automaton" quantum circuits, and use them to study operator growth, information spreading, and local charge relaxation in quantum dynamics with subsystem symmetries, which we define as overlapping symmetries that act on lower-dimensional submanifolds.  With these symmetries, we discover the anomalous subdiffusion of conserved charges; that is, the charges spread slower than diffusion in the dimension of the subsystem symmetry. By studying an effective operator hydrodynamics in the presence of these symmetries, we predict the charge autocorrelator to decay  ($i$) as $\log(t)/\sqrt{t}$ in two dimensions with a conserved $U(1)$ charge along intersecting \emph{lines}, and  ($ii$) as $1/t^{3/4}$  in three spatial dimensions with intersecting \emph{planar} $U(1)$ symmetries. Through large-scale studies of automaton dynamics with these symmetries, we numerically observe charge relaxation that is consistent with these predictions.  In both cases, the spatial charge distribution is distinctly non-Gaussian, and reminiscent of the diffusion of charges along a fractal surface.   We numerically study the onset of quantum chaos in the spreading of local operators under these automaton dynamics, and observe power-law broadening of the ballistically-propagating fronts of evolving operators in two and three dimensions, and the saturation of out-of-time-ordered correlations to values consistent with quantum chaotic behavior.   
\end{abstract}

\maketitle


The dynamics of { interacting}, quantum many-body systems provide a rich source of open problems in theoretical physics. Recent developments include advances in our understanding of quantum thermalization \cite{Deutsch, Srednicki, Rigol} and many-body localization \cite{GMP,BAA,mblarcmp,EhudMBLReview}, which have provided new paradigms for non-equilibrium quantum matter. The study of quantum dynamics  has recently been revolutionized by the study of {random quantum circuits} \cite{Nahum1, PhysRevX.8.021014, Nahum3,  KhemaniVishHuse,  Keyserlingk1, Keyserlingk2, Nahum4}, which have been a source of theoretically tractable problems that shed light on the dynamics of more general quantum systems. 

A richer set of quantum dynamical phenomena arise in the presence of {\it conservation} laws. A global $U(1)$ symmetry leads to the { diffusion} of the conserved charges \cite{KhemaniVishHuse, Keyserlingk1}. Overlapping conservation laws lead to more striking features in quantum dynamics. For example, the conservation of both the total $U(1)$ charge and dipole moment in one spatial dimension can lead to the breaking of ergodicity and localization, and a ``shattering" of the Hilbert space into exponentially many dynamically disconnected sectors  \cite{pai2018localization, Munich, KN}. Most studies of quantum dynamics with or without conservation laws have thus far been limited to systems in one spatial dimension.  

In this work, we extend the study of quantum dynamics in two and three spatial dimensions, by introducing a class of quantum dynamics -- termed \emph{automaton dynamics} -- for which the evolution of various correlation functions of local operators is classically simulable, and which retain certain key features of the dynamics of a chaotic quantum system. 
These dynamics share the property that the {Heisenberg evolution} of a local operator is complex, in a manner that resembles the evolution under a more general, chaotic quantum dynamics; nevertheless, these quantum dynamics are simulable since they do not generate \emph{any} entanglement when acting on a complete set of product states in an appropriate basis.  
We identify key features of the most general automaton dynamics -- in the Heisenberg evolution of local operators, the recurrence times for an initial state, and in the generation of operator entanglement -- that distinguish them from other kinds of classically simulable quantum dynamics, and conjecture that the evolution of local observables under these dynamics can be quantifiably similar to that of a more generic, chaotic quantum system with the same symmetries.   Two examples of automaton circuits, which necessarily generate operator entanglement, have been previously studied in one spatial dimension, and one of these is known to be integrable \cite{Sarang1, GopalakrishnanBahti, log_op_entanglement}.


We apply our understanding of these automaton dynamics to the study of quantum dynamics with {\it subsystem} symmetries -- symmetries that act along overlapping, subdimensional ``manifolds" of the system -- and find a rich and unexpected behavior involving {\it anomalous subdiffusion} of charge, slower not only than diffusion in that physical dimension, but also slower than diffusion in the dimension of the submanifold acted on by the symmetry.  Our motivation to study these dynamics arises from the existence of exotic quantum phases of matter  with immobile, fractionalized excitations (``fracton" phases) \cite{chamon, fracton1, haah, fractonarcmp} in which these symmetries,  such as the conservation of multipole moments of charge \cite{sub},  are emergent properties of these stable {phases} \cite{fracton2}, that severely restrict the dynamics of these exotic, fractionalized excitations.

In this work, we specifically consider Floquet dynamics that conserve $U(1)$ charge along intersecting lines in two and three dimensions, and three dimensional dynamics that conserve $U(1)$ charge along planes, with no additional symmetries; effectively, these are infinite-temperature dynamics in which these subsystem symmetries are exact. For a two dimensional system with line-like symmetries, we find that the local charge relaxes  as $\log(t)/\sqrt{t}$, whereas for a three dimensional system with planar symmetries it relaxes as $1/t^{3/4}$. These results originate from  an analytical explanation in terms of the emergent operator hydrodynamics of the problem, and are consistent with extensive numerical studies of automaton dynamics with these symmetries. Our results are summarized in Table \ref{tab:dynamics_summary}. This anomalous subdiffusion is tied to a distinctly non-Gaussian charge distribution, and reminiscent of classical fractional subdiffusion \cite{METZLER20001,henry2010introduction}.  

\begin{table}[t]
    \begin{tabular}{| c | c | c |} 
      \hline
      \textbf{Subsystem Symmetry} & \multicolumn{2}{|c|}{\textbf{ Charge Decay }}  \\
      \hline
       & {$\mathit{d=2}$} & {$\mathit{d=3}$} \\
       \hline
       \emph{Line Symmetry} & \,\,$\log(t)/\sqrt{t}$\,\, & \,\,$\log^{2}(t)/\sqrt{t}$\,\, \\
       \hline
       \emph{Plane Symmetry} & --- & $1/t^{3/4}$ \\
       \hline
\end{tabular}
        \caption{{\bf Subdiffusive Dynamics with $U(1)$ Subsystem Symmetries} -- Summary of analytical results for the subdiffusive decay of charges in two and three spatial dimensions, in the presence of overlapping line-like and planar $U(1)$ symmetries.  These predictions for line symmetries in $d = 2$ and planar symmetries in $d=3$ are consistent with numerical studies of automaton dynamics with these symmetries.}\label{tab:dynamics_summary}
\end{table}

An efficient numerical simulation of dynamical (both time-ordered and out-of-time-ordered) correlation functions in automaton circuits is possible due to the fact that these correlations may be computed using a Monte Carlo sampling procedure.  This provides a unique example of a quantum dynamics, where we can measure these correlations in two and three spatial dimensions for a highly entangled operator evolution. In two dimensions, we are able to simulate systems with up to $N=396\times 396$ sites and circuits with up to 32000 layers. In three dimensions, we present results for systems with $N=128\times 128\times 128$ sites and circuit depths up to 21000 layers.  Observables other than the conserved charge display apparently ``quantum chaotic" behavior;  out-of-time-ordered correlations of these observables propagate ballistically, and saturate to values that suggest that the structure of these Heisenberg-evolved operators equilibrates in a manner that is consistent with quantum chaos.  The widths of the ballistically-propagating fronts of these operators grow in time as a power-law $t^{\alpha}$ with exponent $\alpha = 0.308(18)$ in two dimensions and $\alpha = 0.220(5)$ in three dimensions.  These should be compared with the predicted values $\alpha = 1/3$ and $\alpha = 0.24$ in two and three dimensions, respectively, for random unitary quantum circuits \cite{PhysRevX.8.021014}.

This paper is structured as follows. In Sec.~\ref{model} we define and present general features of automaton dynamics, and argue that a generic automaton evolution can lead to the complex Heisenberg evolution of local operators, as quantified by the generation of a large degree of ``operator entanglement". For certain operators, we are able to solve for the Heisenberg evolution explicitly.  We then present the numerical algorithm employed for our simulations. Our comparison of automaton evolution to other quantum dynamics is presented in Table \ref{tab:automaton_summary}. In Sec.~\ref{results} we present our results on the anomalous subdiffusion of charges in the presence of various subsystem symmetries. In 
Sec.~\ref{chaos} we study the evolution of out-of-time-ordered correlation functions under these automaton dynamics, and show that the Heisenberg evolution of operators other than the conserved charges appears quantifiably similar to that of a quantum chaotic system. We conclude in Sec.~\ref{conclude} with a discussion of the implications of our results.

\section{Automaton Dynamics} 
\label{model}

We begin by defining the ``automaton" dynamics of a quantum system as a unitary evolution that ($i$) does not generate any entanglement in an appropriate basis of product states, but that ($ii$) leads to the non-trivial evolution of local operators under Heisenberg evolution.  

More precisely, an automaton unitary operator $U$ acting on an appropriate set of orthonormal product states in a $D$-dimensional Hilbert space -- labeled  $|m\rangle$, with $m \in \{0,\,\ldots,D-1\}$ -- simply permutes these states up to a phase factor, i.e.
\begin{align}\label{eq:U_aut}
U\ket{m} = e^{i\theta_{m}}\ket{\pi(m)},
\end{align} 
where $\pi\in S_{D}$ is an element of the permutation group on $D$ elements.   An automaton unitary will generally create entanglement when acting on product states in a different basis.   We refer to Eq. (\ref{eq:U_aut}) as the ``automaton constraint" for the remainder of this section.  

While the evolution of the product states $\{|m\rangle\}$ is simple, the Heisenberg evolution of a local operator
\begin{align}
\mathcal{O} \rightarrow U^{\dagger}\mathcal{O} U
\end{align}
can be complex, in a manner that resembles the Heisenberg evolution of an operator under a more generic, chaotic quantum dynamics; we will quantify this similarity in later sections.  For this reason, automaton unitary evolution is of interest as an example of classically simulable, quantum dynamics that may capture broader features of the dynamics of chaotic quantum systems.  

\begin{table*}[t]
    \begin{tabular}{| c | c | c | c | c |} 
      \hline
    \underline{Features} &    $\begin{array}{c} \textbf{{Integrable}}\\\mathbf{{Dynamics}}\end{array}$ & $\begin{array}{c} \mathbf{{Clifford}}\\\mathbf{{Dynamics}}\end{array}$ & $\begin{array}{c} \mathbf{\color{blue}{Automaton}}\\\mathbf{\color{blue}{Dynamics}}\end{array}$    & $\begin{array}{c} \mathbf{{Quantum\,\,Chaotic}}\\\mathbf{{Dynamics}}\end{array}$\\
      \hline
       $\begin{array}{c} \mathit{Generates \,Volume\text{-}Law}\\\mathit{State \,Entanglement}\end{array}$ & Yes & Yes & $\begin{array}{c} \color{blue}\text{For states in}\\\color{blue}\text{a specific basis }\end{array}$ & Yes\\
       \hline
       $\begin{array}{c} \mathit{Generates\, Volume\text{-}Law}\\\mathit{Operator\,Entanglement}\end{array}$ & $\begin{array}{c} \text{Yes}\end{array}$& No & \color{blue} Yes & Yes \\
       \hline
       $\begin{array}{c} \text{\emph{Recurrence Time }}\\\text{{(product initial state in 1D)}}\end{array}$ & $\mathrm{poly}(\log D)$ & $\log D$ ($\dagger$) & \color{blue}$D$ & $\exp(\lambda D)$\\
       \hline
        \text{\emph{Operator Spreading}} & $\begin{array}{c}\text{Ballistic Growth,}\\\text{Diffusive Broadening}\end{array}$ & $\begin{array}{c}\text{Ballistic Growth,}\\\text{No Broadening}\end{array}$  & $\begin{array}{c}\color{blue}\text{Ballistic Growth,}\\{\color{blue}\text{Power-Law Broadening\,}} \end{array}$ & $\begin{array}{c}\text{Ballistic Growth,}\\\text{Power-Law Broadening}\end{array}$\\
       \hline
	\hline
       $\begin{array}{c}\emph{Operations to simulate }\\
       \emph{one timestep of evolution} \end{array}$ & $\mathrm{poly}(\log D)$ & $\mathrm{poly}(\log D)$   & \color{blue}$D$ & $D^{2}$\\
       \hline
       \emph{Memory Cost} & $D$ ($\ddagger$) & $\log^{2} D$   & \color{blue}$D$ & $D$\\
       \hline
\end{tabular}
        \caption{{\bf Properties of a Generic Automaton Unitary Dynamics}, as compared to Clifford dynamics, a generic integrable dynamics, and the dynamics of a quantum chaotic system, each with a $D$-dimensional Hilbert space.  The typical recurrence time for a state evolving under an integrable dynamics in one dimension grows polynomially in the system size, since each quasiparticle takes a time $O(\log D)$ to return to its initial position; furthermore, ($\dagger$) assumes a translationally invariant, Floquet Clifford circuit \cite{Keyserlingk2}.  Operator spreading, as quantified by out-of-time-ordered correlations, reveals a power-law broadening of the front of an evolving operator in time for automaton circuits in this work, which is close to what has been observed in random unitary dynamics in various spatial dimensions \cite{PhysRevX.8.021014}, unlike what is observed for integrable dynamics \cite{integrable_op_spreading}. The computational cost of simulating one timestep of evolution, by applying $O(\log D)$ gates is indicated, with results for integrable systems from Ref. \cite{Valiant, 1d_efficient}.  The memory overhead is the memory cost to store an evolving state under the indicated dynamics at long times; this is generally $O(D)$ as a product initial state will generally become volume-law entangled under each of these dynamics. For Clifford dynamics, the memory cost is significantly reduced for initial states in the Pauli basis \cite{Gottesman_Knill}. For integrable dynamics ($\ddagger$), the operator entanglement is believed to grow logarithmically in time \cite{log_op_entanglement, Prosen_log}, which reduces the cost of storing an evolving state, though the state may eventually become volume-law entangled.}\label{tab:automaton_summary}
\end{table*}

Before studying the evolution of local operators under automaton dynamics, it is useful to contrast the simulability of these dynamics with integrable dynamics, Clifford unitary evolution, and the dynamics of a chaotic quantum system.  Our comparison is summarized in Table \ref{tab:automaton_summary}.  We note, in particular, that  automaton dynamics are qualitatively the opposite of a Clifford quantum circuit.  In a Clifford unitary dynamics, any Pauli operator (for a spin-(1/2) system, these are the Pauli $X$, $Y$, $Z$ operators, and their products) simply evolves into a {product} of Pauli operators under Heisenberg evolution. While operator evolution is not complex, Clifford unitary operators can generate a high degree of quantum entanglement when acting on any wavefunction. 

An important example of a unitary operator that can generate an automaton quantum circuit -- defined as a product of local, automaton unitary operators that each do not generate entanglement in the same basis -- is a variant of the controllled-controllled NOT gate, a three-qubit gate which we define as
\begin{align}\label{eq:CCNOT_example}
\mathrm{CCNOT}(\theta)_{123} \equiv 1 - \Pi_{12} + \Pi_{12}\,e^{i\theta}X_{3}.
\end{align}
Here $\Pi_{12} = |{\uparrow_{1}\uparrow_{2}}\rangle|\langle{\uparrow_{1}\uparrow_{2}}|$ is the projector onto the up-state of spins $1$ and $2$, while $X_{3}$ is the Pauli $X$ operator acting on spin $3$.  Since this unitary transformation conditionally flips the third qubit based on the states of the first two qubits, it generates no entanglement in the Pauli $Z$ basis. However, the Heisenberg evolution of a local operator is more complex, e.g.  $Z_{3} \rightarrow (1/2)\left[ Z_{3} - Z_{1}Z_{3} - Z_{2}Z_{3} - Z_{1}Z_{2}Z_{3}\right]$. 
We note that ($i$) $\mathrm{CCNOT}(\theta=0)$ is the Toffoli gate, which is universal for classical, \emph{reversible} computing \cite{Nielsen_Chuang}, as any invertible Boolean function can be constructed using this gate and ancilla bits, and that ($ii$) the Toffoli gate, and the single-qubit Hadamard gate, define a universal gate set \cite{toffoli_universal}.

\subsection{Growth of the Operator Entanglement}
We now study the growth of the complexity of an operator under the Heisenberg evolution by an automaton quantum circuit.  For simplicity of presentation, and without loss of generality, we restrict our attention to a spin-(1/2) system. Under Heisenberg evolution by a chaotic unitary dynamics, a local Pauli operator $\mO$ in a chaotic quantum system will evolve into sums of arbitrary products of Pauli operators as 
\begin{eqnarray}\label{eq:op_evolution}
\mathcal{O}(t) = \sum_{\mS} a_\mathcal{S}(t) \mathcal{S},
\end{eqnarray}
where the sum is over all operators $\{\mS\}$ that are products of Pauli operators on distinct sites in the system.  Since $\mO(t)^{2} = 1$, the real coefficients $a_\mathcal{S}(t)$ satisfy the condition
$\sum_{\mathcal{S}} a_{\mS}(t)^2 = 1$, for all times $t$.  

We now consider an automaton unitary operator $U(t)$ that generates no entanglement when acting on states in the  Pauli $Z$-basis.  In other words, a projection operator $|m\rangle\langle m|$ in this basis evolves as $U(t) |m\rangle\langle m| U^{\dagger}(t) = |\pi(m)\rangle\langle \pi(m)|$.  A simple consequence of this is that a Pauli $Z$ operator will only evolve into a sum of products of other Pauli $Z$ operators under Heisenberg evolution.
 
 While this restricted evolution is a special feature of the automaton dynamics, we now show that under a generic automaton evolution, a Pauli $Z$ operator can grow to develop a weight on any possible operator within this sub-space.  More precisely, we demonstrate that the ``operator entanglement" -- which quantifies the complexity of the growing operator $U^{\dagger}(t)\,Z\,U(t)$ -- can develop volume-law scaling under an automaton evolution. 

To proceed, we let $\mS_{n}$ denote a product of $Z$ operators at sites corresponding to the binary representation of the integer $n \in \{0,\ldots, D-1\}$. For example, the operator $\mS_{5} = Z_{1}Z_{3}$ (since the binary representation of $5$ is $10100\cdots$).  These operators evolve as 
\begin{align}
\mS_{n}(t) = U(t)\mS_{n}\,U(t)^{\dagger} = \sum_{m=0}^{D-1}a_{nm}(t) \,\mS_{m}.
\end{align} 
The ``automaton constraint"  permits us to determine in  Appendix \ref{appendix:op_spreading} that the coefficients $a_{nm}(t)$ are exactly
\begin{align}\label{eq:anm_text}
a_{nm}(t) = \frac{1}{D}\sum_{k=0}^{D-1}(-1)^{\boldsymbol{\pi}(k)\cdot\boldsymbol{m} + \boldsymbol{k}\cdot\boldsymbol{n}}.
\end{align}
Here, $\boldsymbol{\pi}(k)$ is the binary vector representation of the integer $\pi(k)$, and $\boldsymbol{\pi}(k)\cdot\boldsymbol{m}$ and $\boldsymbol{k}\cdot \boldsymbol{n}$ denote the dot product of the appropriate vectors. As we show in Appendix \ref{appendix:op_spreading}, the fact that the matrix $a_{nm}(t)$ is {unitary} is a consequence of the unitarity of the original dynamics.  

 To quantify the growth of the complexity of the Pauli operator $\mS_{n}(t)$, it is natural to study its ``operator entanglement", which we define as the entanglement of the evolving wavefunction 
\begin{align}\label{eq:op_state}
|{{n}(t)}\rangle \equiv \sum_{n=0}^{D-1} a_{nm}(t)\,|{m}\rangle,
\end{align}
where $|{m}\rangle$ denotes the state of a fictitious $N$-spin system, corresponding to the binary representation of the integer $m$, so that a ``$1$" (``$0$") in this binary representation corresponds to a down (up) spin; for example, for $m = 5$, we have the state $|{m}\rangle = |\downarrow\uparrow\downarrow\uparrow\uparrow\uparrow\cdots\rangle$.   We observe that the unitary operator $W(t)$, whose matrix elements are $\langle{m}\,|\,W(t)\,|\,{n}\rangle \equiv a_{nm}(t)$, can be written as 
\begin{align}
W(t) = H^{\otimes N} U(t) H^{\otimes N},
\end{align}
where $H$ is the single qubit Hadamard gate, which acts on Pauli operators as $HXH = Z$, $HZH = X$.  In other words, the operator entanglement evolves according to the \emph{same} automaton circuit in a rotated basis; in this basis, a generic automaton unitary operator will eventually produce a high degree of entanglement.  

Instead of demonstrating this explicitly by studying the growth of operator entanglement for a particular automaton evolution, we quantify the operator entanglement generated by a random automaton unitary operator.  We bi-partition the state $|\boldsymbol{n}(t)\rangle$ into an $A$ and $B$ subsystem, of Hilbert space dimension $D_{A}$ and $D_{B} = D/D_{A}$, respectively.  The reduced density matrix for the $A$ system is defined as
\begin{align}
\rho_{A}(n) \equiv \Tr_{B}\,|{\boldsymbol{n}(t)}\rangle \langle {\boldsymbol{n}(t)}|.
\end{align}
We determine in Appendix \ref{appendix:op_entanglement}, that the purity of this density matrix for a {random} automaton unitary, corresponding to a random permutation $\pi\in S_{D}$, is 
\begin{align}
\overline{\,\Tr\,\rho_{A}(n)^{2}\,} = \left\{ \begin{array}{cc} 1 & (n=0)\\
&\\
\displaystyle{D_{A}^{-1} + D_{B}^{-1} - D^{-1}} & (n\ne 0)
\end{array}\right.,
\end{align}
where the line $\overline{\cdots}$ denotes an average over automaton unitary operators that do not generate entanglement in the Pauli $Z$ basis.  For any finite subsystem $A$, taking the thermodynamic limit $D\rightarrow\infty$, $D_{A}/D\rightarrow 0$ yields $\overline{\,\Tr\,\rho_{A}(n)^{2}\,} = D_{A}^{-1}$ when $n\ne 0$, which is the \emph{maximally-entangled} value of the density matrix.  We observe that, $\Tr\,\rho_{A}(0)^{2} = 1$ since the state $|\boldsymbol{n} = 0\rangle$ is an eigenstate of $W(t)$, as this state corresponds to the trivial evolution of the {identity} operator. 


We now argue that time evolution must generate a high degree of operator entanglement for an initial operator that is a product of Pauli $X$ operators. Instead of attempting to solve for their Heisenberg evolution, we observe that any product of these operators -- which we label $\mO$ -- implements a permutation $\tau\in S_{D}$ when acting on states in the $Z$ basis.  As a result, 
\begin{align}
U(t)^{\dagger}\mO U(t) = \sum_{m=0}^{D-1} e^{i(\theta_{m}-\theta_{\tau\pi(m)})}\,|\pi^{-1}\tau\pi(m)\,\rangle\langle{m}|
\end{align}
if the automaton unitary $U(t)$ generates no entanglement in the $Z$ basis.  If we let $\theta_{n} = 0$ for all $n$, then the Heisenberg-evolved operator $\mO(t) = U(t)^{\dagger}\mO U(t)$ can only act as an element of its \emph{conjugacy class}, i.e. as an element of the form $\sigma^{-1}\tau\sigma$, for some $\sigma\in S_{D}$.  Conjugacy classes of the permutation group are labeled by their cycle type, defined as the lengths of all cycles in an element of that class. Since $\mO$ squares to the identity and is traceless, the permutation $\tau$ consists of $D/2$ independent transpositions. The size of this conjugacy class is 
 \begin{align}
\frac{D!}{2^{D/2}(D/2)!} \overset{D\rightarrow\infty}{\sim} D^{({D}/{2}) \left[1 + O(\log^{-1} D)\right]},
\end{align}
which is the number of possible operators that $\mO(t)$ can evolve into.  As this is much larger than the $O(D^{2})$ operators which are simple products of Pauli operators in the system, we conclude that $\mO(t)$ must exhibit a high degree of operator entanglement, for a randomly chosen automaton unitary $U(t)$.  If the phases in this automaton unitary  are non-zero ($\theta_{n} \ne 0$), then $\mO$ no longer evolves into a countable set of operators.  Nevertheless, the above argument for the growth of the operator entanglement remains valid.   

\subsection{The Recurrence Time and Quantum Chaos}\label{sec:q_chaos_sec}
We compare the recurrence times -- when a given initial state returns to itself, so that $|\langle \psi |U(t)|\psi\rangle| \sim O(1)$ -- for a typical automaton circuit, with that of a chaotic quantum system.   For a chaotic quantum system, the Poincare recurrence time scales exponentially in the \emph{Hilbert space dimension} of the system $t_{\mathrm{rec}} \sim \exp(\lambda D)$.  In contrast, let $U$ be a random automaton operator that generates a single timestep of a Floquet unitary operator $U(t) = U^{t}$, so that $U$ is proportional to a random permutation of product states in a $D$-dimensional Hilbert space.  The behavior of the recurrence time for $U(t)$ can be varied.  First, the recurrence time for a random product state for which $U$ generates no entanglement will grow as
\begin{align}
t_{\mathrm{prod}} \sim D/2
\end{align}
as we show in Appendix \ref{appendix:permutation_group}. In contrast, the recurrence time for a random state (not necessarily a product state) will be much larger, as this corresponds to the order of a random element of the permutation group $\pi \in S_{D}$ whose asymptotic form as $D\rightarrow \infty$ \cite{Rand_perm_order} gives  
\begin{align}
t_{\mathrm{rand}} \overset{D\rightarrow\infty}{\sim} \exp\left[\lambda\sqrt{D/\log{D}}\right],
\end{align}
where $\lambda$ is an $O(1)$ constant.  

Finally, to compare the operator evolution under an automaton dynamics with the evolution in a chaotic quantum system, we consider the \emph{out-of-time-ordered} correlation function 
\begin{align}\label{eq:OTOC_main_text}
F(\boldsymbol{r}, t) \equiv \frac{1}{D}\Tr [\mO_{\boldsymbol{r}}(t)\mO'_{0}(0)\mO_{\boldsymbol{r}}(t)\mO'_{0}(0)],
\end{align}
where $\mO_{\boldsymbol{r}}(0)$ and $\mO'_{0}(0)$ are two initially local operators, at the indicated positions.  The out-of-time-ordered correlation function probes the structure of the evolving operator  $\mO_{\boldsymbol{r}}(t)$, which we may expand as a sum of products of Pauli operators as in Eq. (\ref{eq:op_evolution}).  For $\mO'_0(0)$ a Pauli string, we observe that  
\begin{align}
F(\boldsymbol{r}, t) = 1 - 2 \sum_{\{\mS, \mO'_{0}\} = 0} a_{\mS}(t)^{2}.
\end{align}
The quantity $\sum_{\{\mS, \mO'_{0}\} = 0} a_{\mS}(t)^{2}$ is just the probability that $\mO_{\boldsymbol{r}}(t)$ has weight on a Pauli operator that anti-commutes with $\mO'$.  For a chaotic quantum system, a typical operator  grows {ballistically} under Heisenberg evolution, with a \emph{butterfly velocity} $v_{B}$, and within this growing region, the operator is equally likely to develop a weight on Pauli operators that commute or anti-commute with $\mO'_{0}$ so that $\sum_{\{\mS, \mO'_{0}\} = 0} a_{\mS}(t)^{2} \rightarrow (1/2)$ and
\begin{align}
F(\boldsymbol{r}, t)\sim \left\{\begin{array}{cc}
0 &  \,\,\,\,v_{B}t\gg |\boldsymbol{r}|\\
&\\
1 & \,\,\,\,v_{B}t\ll |\boldsymbol{r}|
\end{array}\right..
\end{align}
Therefore, OTOC ($i$) has a ballistically growing front, and ($ii$) within this region the OTOC asymptotically vanishes due to the equilibration of the structure of $\mO_{\rB}(t)$.  We will confirm that these features, which hold for chaotic quantum dynamics, are also present for automaton unitary circuits in our numerical studies.  We note that as before, one can show using Eq. (\ref{eq:anm_text}) that for a \emph{random} automaton dynamics $U$, that
\begin{align}
\overline{\,\Tr [U^{\dagger}Z_{0}UZ_{0}U^{\dagger}Z_{0}UZ_{0}]\,} = 0,
\end{align}
where $\overline{\cdots}$ again denotes an average over the choice of automaton circuits. 

Finally, the front of the OTOC is believed to broaden as a power-law in time $\sim t^{\alpha}$ in a chaotic quantum system, where $\alpha$ is a dimension-dependent exponent. In $d=1$, the ends of an operator perform a biased random walk, leading to the exponent $\alpha = 1/2$ \cite{PhysRevX.8.021014, Keyserlingk2}, while in higher dimensions, these exponents are related to probability distributions for classical stochastic growth processes \cite{PhysRevX.8.021014}.  We verify that a power-law broadening of the operator front occurs for the OTOC in the automaton circuits we consider.
 

\subsection{Simulating Automaton Evolution}

We now discuss how dynamical correlation functions may be efficiently calculated for an automaton evolution, using classical
Monte Carlo techniques.  Let $U(t)$ be an automaton unitary evolution that generates no entanglement in the Pauli $Z$ basis.  We wish to determine the weight $a_{\mS}(t) = (1/D)\,\Tr[\mO_{\boldsymbol{r}}(t)\mS]$ of an evolving Pauli operator $\mO_{\boldsymbol{r}}(t) \equiv U(t)^{\dagger}\mO_{\boldsymbol{r}} U(t)$ on each basis string $\mathcal{S}$.  We observe that
\begin{eqnarray}
\hspace{-2mm} a_\mathcal{S}(t) &=& \frac{1}{D}\sum_{n}\langle n|\mO_{\boldsymbol{r}}(t)\mathcal{S}|n\rangle  \nonumber\\
&=& \frac{1}{D}\sum_{n,m} e^{i(\theta_{m}-\theta_{n})} \langle n(t)| \mO_{\boldsymbol{r}} | m(t) \rangle\, \langle{m}|\mathcal{S}|{n}\rangle.
\end{eqnarray}
This quantity can be easily calculated
using classical methods since $\ket{n(t)}=U(t)\ket{n}$ remains a product state for any product state 
$\ket{n}$ in the computational basis. A further speedup is obtained because for strings diagonal in the computational basis (e.g. products of Pauli $Z$ operators), $\bra{m}\mathcal{S}\ket{n} = \pm \delta_{mn}$, so one only needs to sample over $|n\rangle$, whereas for a string off-diagonal in the computational basis (e.g. $X$) one only needs to sample over $|n\rangle$ and $|m\rangle = X_i |n\rangle$, instead of needing to sample over $|n\rangle$ and $|m\rangle$ independently. This idea of classically
sampling a quantum wave function is standard in variational Monte Carlo methods.
Here, we extend this idea to sampling a class of entangled quantum operators.  The coefficients of the basis strings in an expansion of the highly entangled `variational quantum operator' are generated by the action of the unitary circuit on an initially local operator.  Using this method we can study the unitary time evolution of very
large quantum systems.  In this paper, we look at 2D circuits with up to $396^2$ sites and circuit depths of up to 32000 layers, as well as 3D circuits with up to $128^3$ sites and circuit depths of up to 21000 layers.

\section{Results on Automaton Circuits with Subsystem Symmetries}
\label{results}
In this section, we study automaton quantum dynamics that possess an extensive set of intersecting, global symmetries.  In this case, the conserved charges cannot move in isolation, but engage in a complex, correlated motion.  We demonstrate through analytical arguments and numerical studies of these quantum dynamics, that the conserved charges in systems with overlapping ($i$) line-like $U(1)$ symmetries in two spatial dimensions, and ($ii$) planar $U(1)$ symmetries in three spatial dimensions, evolve {subdiffusively}.

\begin{figure}[t]
\includegraphics[scale=0.50]{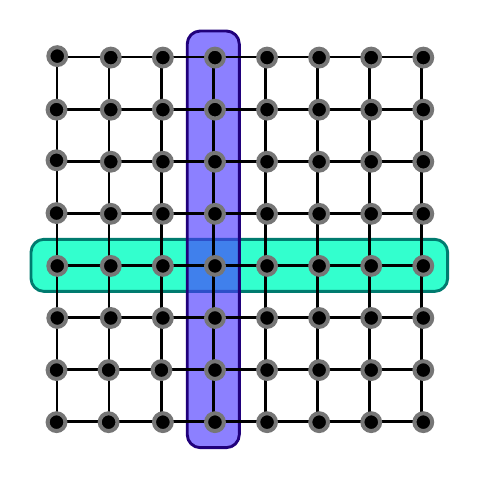}
\hspace{5mm} \includegraphics[scale=0.55]{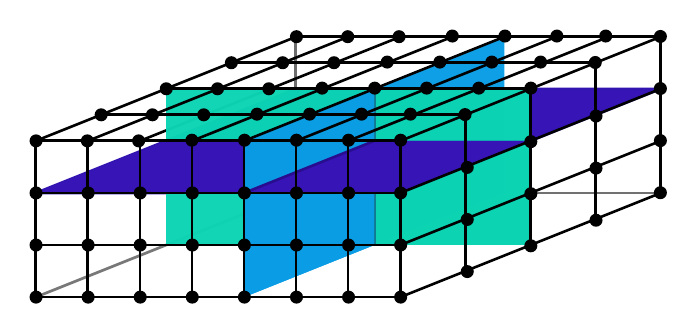}
\includegraphics[scale=0.70]{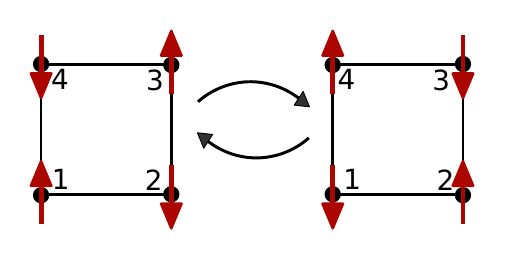}
\includegraphics[scale=0.60]{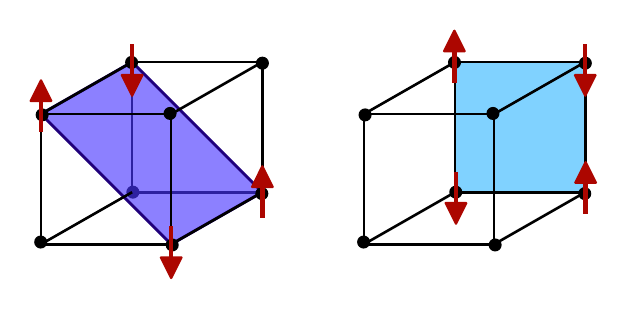}
\caption{{\bf Line-like \& Planar Subsystem Symmetries.}  {\it Top}: We study 2D square lattices with linear subsystem symmetry
where charge is conserved on each row and column of the lattice. We also look at
3D cubic lattices with planar subsystem symmetry. In this case, each plane
contains a conserved $U(1)$ charge. {\it Bottom}: The 4-site unitary gates act
non-trivially only on the above two configurations. In three dimensions, such a
4-site gate can act along any plane of the cube which intersects 4 spins.}
\label{fig1}
\end{figure}


Before proceeding to a discussion of subsystem symmetries, we note that if the unitary circuit of interest possesses a continuous, global symmetry, then there exists a
set of conserved charges, which constrain the Heisenberg evolution of operators. For example, for a unitary evolution that preserves the total $Z$ spin -- so that $\sum_{\boldsymbol{r}} Z_{\boldsymbol{r}}$ is a conserved operator -- the coefficients appearing in the expansion of any operator $\mO(t)$ satisfy the constraint
\begin{eqnarray}
\frac{1}{D}\sum_{\boldsymbol{r}} \Tr[Z_{\boldsymbol{r}} \mO(t)] = \sum_{\boldsymbol{r}} a_{Z_{\boldsymbol{r}}}(t) = \mathrm{const.}
\end{eqnarray}
That is, the existence of a symmetry implies there exists a set of conserved
strings which consist of a single Pauli operator. These are known as the
conserved charges. By dividing an operator into a conserved and non-conserved
part, one can gain a full understanding of the operator dynamics. It will be
important to compare to the known results of unitary circuits with a global
$U(1)$ symmetry. For such a circuit, the support of the non-conserved
operators grows ballistically with time, while the conserved operators
essentially perform a random walk. This leads to diffusive spreading of the
conserved $U(1)$ charge. In turn, the weights of the conserved operator strings
$a_{\mathcal{S}}$ take the form of a Gaussian   
\begin{eqnarray}
\frac{1}{D}\Tr[Z_{\boldsymbol{r}}(t)Z_{0}] \sim {e^{ -({r^2}/{D_{0}t})}}/{(D_{0} t)^{d/2}},
\end{eqnarray} 
where $D_{0}$ is the diffusion constant, and $d$ is the spatial dimension.

\subsection{Dynamics with Line-like Subsystem Symmetries in Two Dimensions}\label{subsec:2D}

We now consider quantum dynamics with \emph{subsystem} $U(1)$ symmetries.  In two spatial dimensions, the simplest such symmetry corresponds to overlapping $U(1)$ charges along intersecting lines.  On the square lattice, we may consider dynamics that preserve the line charges
\begin{eqnarray}
C_x = \sum_y {Z}_{x,y} \,\,\,\,\, \text{and} \,\,\,\,\, C_y = \sum_x {Z}_{x,y},
\end{eqnarray}
which are the total $Z$ spin on each row and column of the square lattice, respectively. 

Consider the case where a single charge is placed at the origin $(0,0)$ on the
lattice, so that $C_{x}=\delta_{x,0}$ and $C_{y}=\delta_{y,0}$. This charge can \emph{only} move while preserving the subsystem symmetries if we also allow for the creation of
new charges.   
In particular, the 2D charge can move along a row or column if it
emits a charge \emph{dipole} in the direction perpendicular to its motion. We can implement
such motion via an automaton unitary gate which preserves the subsystem symmetries.

\begin{figure}[t]
\includegraphics[scale=0.47]{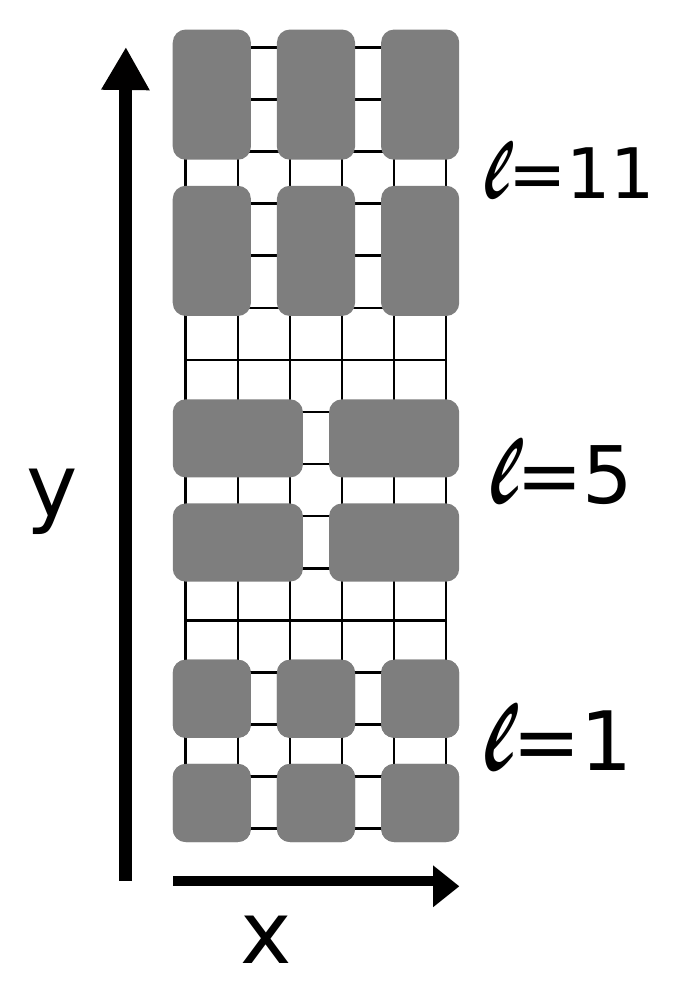}
\hspace{2mm}\includegraphics[scale=0.49]{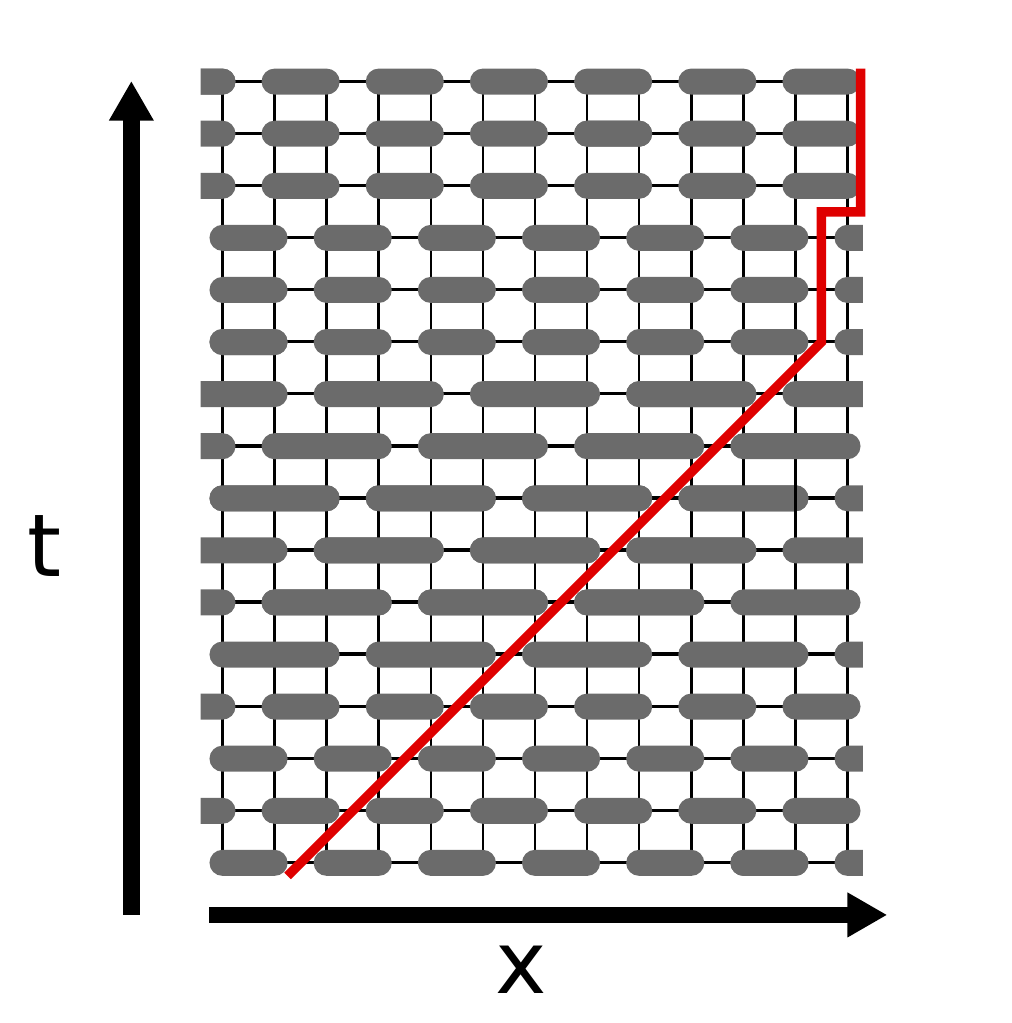}
\caption{{\bf One Timestep of the 2D Automaton Circuit.} {\it Left:} The three types of gates which are applied to the 2D circuit. The displayed gates act on layers $\ell=1,5$ and $11$.  For the next-nearest-neighbor rectangular gates, the unitary acts trivially on the middle two sites of the rectangle. {\it Right:} A single period of the circuit consists of 16 layers, applied as shown in the spacetime cross section. The red line indicate the causal cone for the first site. 
}
\label{circuit}
\end{figure}
In two dimensions on the square lattice, there are no two-site gates which preserve the line-like subsystem
symmetries. The smallest such gate must act on four sites that lie at the corners of a rectangle.
One such gate which preserves the subsystem symmetries is given by
the ``plaquette flip" unitary operator
\begin{eqnarray}\label{eq:U_flip_main_text}
U_{\mathrm{flip}} &=& (1-P) + P X_1 X_2 X_3 X_4,
\end{eqnarray}
where $P$ is the projection operator
\begin{align}
P \equiv \frac{1}{8}(1-Z_1 Z_2)(1-Z_2 Z_3)(1- Z_3 Z_4) \label{eqn2B2}
\end{align}
with the indices $\{1,2,3,4\}$, as labeled as in Fig.~\ref{fig1}. The projection operator
$P$ projects into the subspace spanned by the two states shown in
Fig.~\ref{fig1}. The gate $U$ then flips between the two
states in this subspace and acts trivially on all other states in the four site
region. This automaton unitary operator generates operator entanglement; for example, we observe that
\begin{align}
&U_{\mathrm{flip}}^{\dagger}  Z_{1} U_{\mathrm{flip}} = \frac{3}{4}Z_{1} + \frac{1}{4} \Big[Z_{2} +
Z_{3} - Z_{4}\Big]  \nonumber \\
& +\frac{1}{4} \Big[ Z_{1}Z_{2}Z_{4} + 
 Z_{1}Z_{3}Z_{4} - Z_{1}Z_{2}Z_{3} - Z_{2}Z_{3}Z_{4}  \Big].
\label{ztime}
\end{align} 
We note that the most general unitary operator acting on an elementary plaquette that preserves the line symmetries, is one that performs an arbitrary rotation within the two-dimensional sub-space spanned by the two states shown in Fig. \ref{fig1} -- consisting of the states where each pair of neighboring spins is anti-aligned -- plus any phase gate that is diagonal in the Pauli $Z$ basis. 

We can therefore construct a circuit model which is composed only of these four-site unitary gates. In two dimensions, in addition to acting such gates on the fundamental plaquettes of the square lattice, we also include next-nearest-neighbor gates. In this case, we apply the same four-site unitary to the four corner sites of the rectangular plaquettes shown in Fig.~\ref{circuit}. This is done to improve the ergodicity of the model. We then act the three gate types sequentially on all plaquette coverings of the square lattice in the pattern shown in Fig.~\ref{circuit}. Therefore one periodic timestep of our circuit consists of 16 circuit `layers'.

\subsubsection{Numerical Study}

\begin{figure}[t]
\includegraphics[scale=0.35]{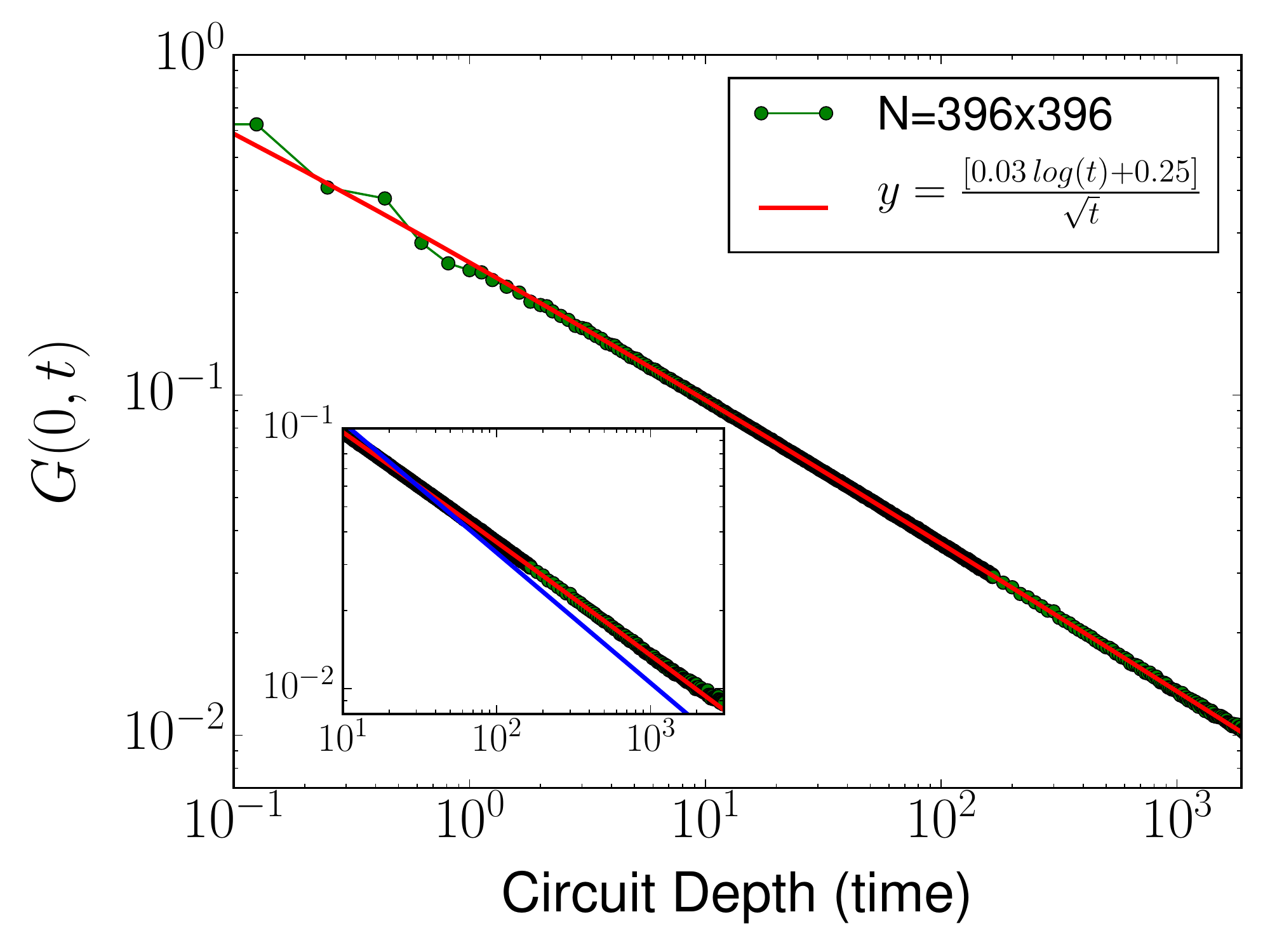}
\caption{{\bf Charge Subdiffusion in 2D} -- The autocorrelation function $G(0,t)$ for the 2D automaton circuit with line symmetries, which decays like
$ {\log(t)}/{\sqrt{t}}$. {\it Inset:} Comparison of $\log(t)/\sqrt{t}$ and
$1/\sqrt{t}$ decay, showing a clear deviation from normal diffusion.}
\label{fig3}
\end{figure}

We start by numerically studying the 2D automaton circuit with linear subsystem symmetries, where we apply the 4-site unitary gates defined in Eq. (\ref{eqn2B2}) on all plaquettes of the square lattice, in the configuration shown in Fig. \ref{circuit}.  We wish to determine the evolution of the correlation function
\begin{align}
G_{\boldsymbol{r}, t}= \frac{1}{D}\Tr[Z_{\boldsymbol{r}}(t)Z_0(0)].
\end{align}
We begin by numerically studying the auto-correlation function when $\boldsymbol{r} = 0$. The results are shown
in Fig.~\ref{fig3}. While one may naively expect the same form for this
correlation function as in the 1D case, since the conserved charges of the
$Z(t)$ operator are constrained to move along the 1D rows and columns of the
lattice, we find numerically that
\begin{eqnarray}
G_{0, t} \sim \frac{a\log(t)+b}{\sqrt{t}}.
\end{eqnarray}
This is in sharp contrast to the 1D and 2D case with a global symmetry where
we expect $G(t) \sim t^{-d/2}$.  We note that it is only by going to very large system sizes and large
circuit depths through the automaton unitary evolution, that we can resolve this logarithmic term in the correlation
function.

Furthermore, we numerically determine the full space-time correlation function $G(\rB,t)$.  Again, in systems with a global conserved charge,
we expect this quantity to show Gaussian behavior.  Our results, shown in
Fig.~\ref{fig4}, vary dramatically from this expectation. In particular, we see
a distinct cusp shape in the spatial correlation function near the $t=0$
location of the charge.  

This anomalous motion of the charge can be understood from the simple time
evolution of the $Z$ operator given in Eq.~\ref{ztime}. Notice that a charge cannot
move freely in this model, but instead can only move by also emitting a dipole
in the direction perpendicular to the motion. This means that motion of
conserved charges in this model necessarily requires the creation of negative
charges. Evidently, the creation of these dipoles results in the creation of a
medium which slows the diffusion of the conserved charge.  We note that this type of anomalous
diffusion can be considered a purely quantum phenomenon, since these 
negative charges are associated with the sign of the operator wave function. 

\subsubsection{Analytical Study of the Charge Subdiffusion}
We can gain some analytic understanding of these dynamics by observing that the overlap of $Z_{0}$ with the conserved line charge is $ \Tr[Z_{0}C_{x}] = D\delta_{x,0}$ and $ \Tr[Z_{0}C_{y}] = D\delta_{y,0}$.  Therefore, for any dynamics that conserves these charges, the dynamical correlation function $G(\rB, t)$ satisfies the constraint that
\begin{align}
\sum_{r_{x}}G_{\rB, t} = \delta_{r_{y},0} \, , \hspace{.4in} \sum_{r_{y}}G_{\rB, t} = \delta_{r_{x},0}. 
\end{align}  
From this, we may construct the simplest evolution equation for $G_{\rB,t}$, after a single timestep of a discrete-time unitary evolution that preserves the line charges $C_{x}$, and $C_{y}$.  Assuming that the dynamics are invariant under four-fold rotations on the square lattice, the simplest difference equation for the evolution of the dynamical correlations is given by
\begin{align} \label{eq:G_finite_diff_2D}
G_{\boldsymbol{r}, t+1}  & = (1-2\lambda)G_{\rB,t}\\
&+ \lambda\left[G_{\rB+x, t} + G_{\rB-x, t} + G_{\rB+y, t} + G_{\rB-y, t}\right]\nonumber\\
& - \frac{\lambda}{2}\left[G_{\rB+x+y, t} + G_{\rB+x-y, t} + G_{\rB-x+y, t} + G_{\rB-x-y, t}\right]\nonumber
\end{align}
where $\lambda$ is a free parameter of the evolution. We note that this is, in fact, an accurate description of the first timestep of the automaton unitary evolution considered in our numerical study with nearest-neighbor gates only (if we take $\lambda = 1/8$); however, Eq. (\ref{eq:G_finite_diff_2D}) neglects ``backflow" effects -- that non-conserved operators appearing in the Heisenberg evolution of $Z_{\rB}(t)$ can develop a non-negligible weight on the operator $Z_{0}$ at sufficiently long times -- which are present in the automaton dynamics that we numerically simulate.  

\begin{figure}[t]
\includegraphics[scale=0.35]{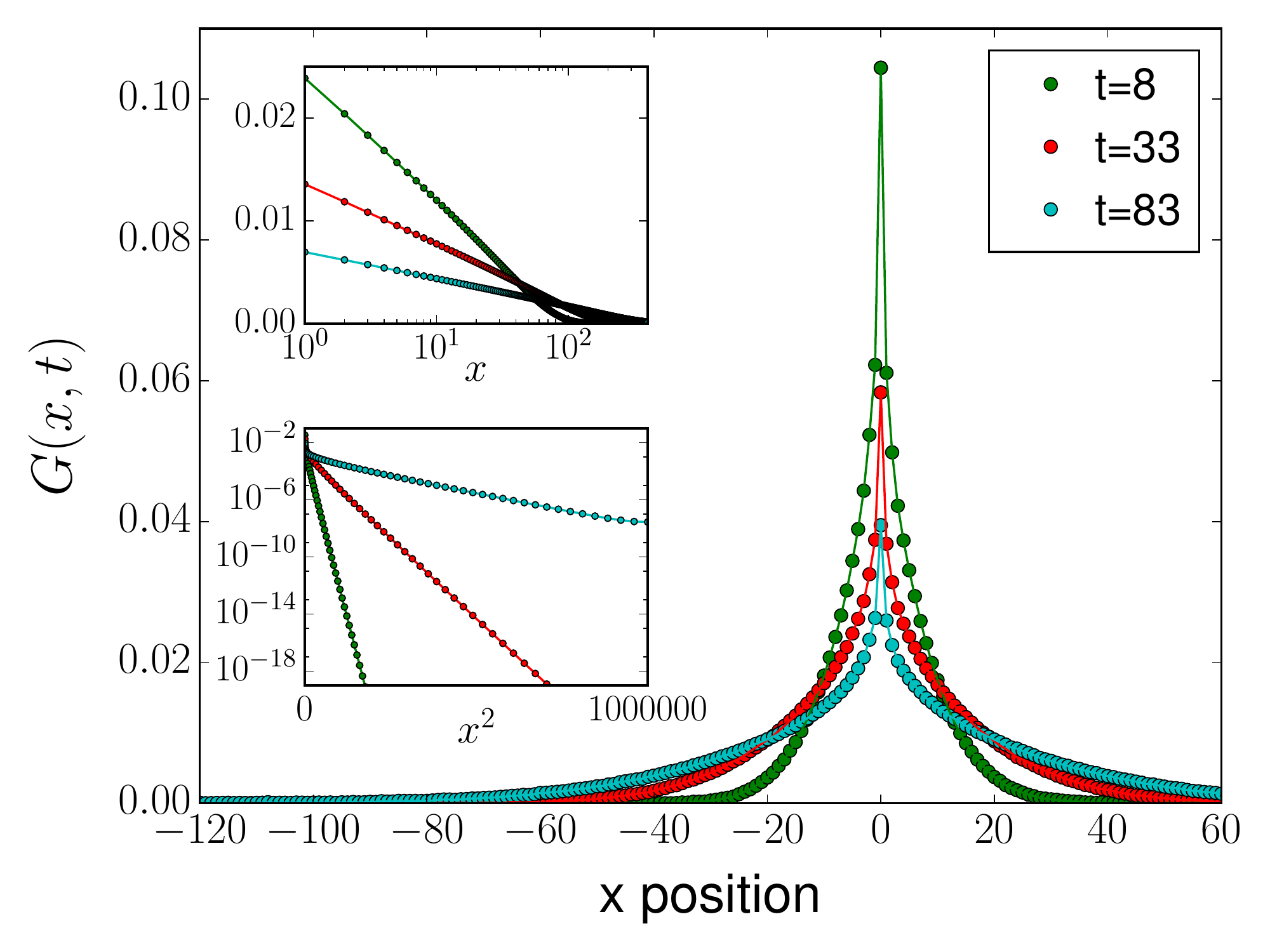}
\caption{{\bf Spatial Charge Distribution in 2D} -- The simulated evolution, with $N=396^2$ sites, of $G_{\rB,t}$ along the line $\rB = (x, 0)$, which shows
a dramatic deviation from the usual Gaussian behavior of normal diffusion.
Notice the distinctive cusp near $x=0$  which is a sign of fractional
subdiffusion. {\it Insets:} The simulated evolution of the 2D difference equation (\ref{eq:G_finite_diff_2D}). For $x \lesssim \sqrt{t}$, $G(x,t)\sim
\log(t/x^2)/\sqrt{t}$. For $x\gg\sqrt{t}$, we have $G(x,t) \sim
e^{-x^2/\sqrt{t}}$ }
\label{fig4}
\end{figure}

Taking the continuum limit in time, and going to momentum space by defining $G(\kB, t) = \frac{1}{\sqrt{N}} \sum_{\rB} \exp(i \kB \cdot \rB) G(\rB, t)$, we find that Eq. (\ref{eq:G_finite_diff_2D}) becomes
\begin{equation}
\frac{d}{dt}G(\kB,t) = - f(\kB) G(\kB,t), \label{kspace2d}
\end{equation}
where $f(\kB) = 8\lambda \sin^2(k_x/2) \sin^2(k_y/2)$. 
We may analytically determine $G(|\rB| = 0,t)$ by taking the inverse Fourier transform, which may be performed to give $G(0,t) = \,_{\frac{1}{2}}F_{\frac{1}{2}}(1; 1; -8\lambda t)$, where $\,_{a}F_{b}(p;q;z)$ is the generalized hypergeometric function. The long time expansion of this expression yields
\begin{equation}\label{eq:long_time_G0}
G(0,t) \overset{\lambda t\rightarrow\infty}{\sim} \frac{\log( \lambda t) + O(1) }{\sqrt{\lambda t}},
\end{equation}
as we show in Appendix \ref{appendix:1d_subsys_dynamics}; this agrees with our numerical study of the automaton dynamics.  We note that the same result may be obtained by taking the continuum limit of Eq. (\ref{eq:G_finite_diff_2D}) in both space and time, to obtain the partial differential equation
\begin{eqnarray}\label{eq:double_laplacian}
\partial_t G(\rB,t) = -\frac{\lambda}{2}\partial^2_x \partial^2_y G(\rB,t).
\end{eqnarray}
By performing a Fourier transform, we may solve for $G(\rB, t)$ in terms of special functions, whose expansion at long times yields the same result as in Eq. (\ref{eq:long_time_G0}).  

We can also numerically simulate the finite difference equation
Eq.~(\ref{eq:G_finite_diff_2D}). The solution $G(\rB,t)$ along the line $\rB = (x,0)$ is plotted in
Fig.~\ref{fig4}. The insets show the behavior in the two limits $x<<\sqrt{t}$
and $x>>\sqrt{t}$, at fixed times $t_0$. We find that this is consistent with 
\begin{eqnarray}
G(x,t) \sim \left\{\begin{array}{cc}\displaystyle
\frac{\log(t/(x+x_0)^2)}{\sqrt{t}} &(x \ll \sqrt{t}) \\
& \\
\displaystyle e^{-x^2/t} & (x \gg \sqrt{t})
\end{array}\right. ,
\end{eqnarray}
where $x_0$ is a constant which we can fit numerically.

Finally, it is interesting to note that, by virtue of conserving charge along lines, this circuit also preserves the dipole moment in both the $x$ and $y$ directions. In Ref.\cite{pai2018localization} it was suggested that dipole conserving circuits in two dimensions should display localization of charge. In contrast, our results here establish that charge does spread, albeit subdiffusively, with the equation for charge spreading containing a term that involves the {\it square} of the Laplace operator. This suggests that the hydrodynamics for dipole conserving circuits proposed in \cite{pai2018localization} cannot be complete, and furthermore suggests that the missing ingredient may be a Laplacian squared term. 

Finally, we may consider the dynamics of a conserved charge in a three-dimensional system with intersecting line-like symmetries along three orthogonal directions, which are given by the total $Z$ spin along each of these directions.  As we demonstrate in Appendix \ref{appendix:line_symm_3d}, the three-dimensional generalization of Eq. (\ref{eq:double_laplacian}) -- given by $\partial_{t}G(\rB, t) = \lambda\,\partial_{x}^{2}\partial_{y}^{2}\partial_{z}^{2}G(\rB,t)$ -- may be solved in terms of a special function, whose asymptotic form when $\lambda t \gg x^{2}y^{2}z^{2}$ yields the result that
\begin{align}
    G(\rB, t) \,{\sim}\, \frac{\log^{2}(\lambda t) + O(\log(\lambda t))}{\sqrt{\lambda t}}\hspace{.23in} ({\lambda t\gg x^{2}y^{2}z^{2}}).
\end{align}
We have not confirmed this through numerical simulations of an automaton circuit, and we leave a detailed study of automaton dynamics with line-like symmetries in three spatial dimensions to future work. 

\subsection{Dynamics with Planar Symmetries in Three Dimensions}
We can also consider the case of intersecting planar symmetries on the 3D cubic
lattice. In this case, a $U(1)$ charge must be conserved on all $xy$, $xz$ and $yz$
planes.  For spin-(1/2) degrees of freedom on the sites of a cubic lattice, we may define these charges as the total $Z$ spin on each plane, 
\begin{eqnarray}
C_x = \sum_{y,z} \hat{Z}_{xyz} \hspace{1.5mm} , \hspace{1mm} C_y = \sum_{x,z}
\hat{Z}_{xyz}, \hspace{1.5mm} 
\hspace{1mm} C_z = \sum_{x,y} \hat{Z}_{xyz}.
\end{eqnarray}
Automaton dynamics that respect these symmetries, along with the symmetries of the cubic lattice
may be implemented using four-site unitary gates. In particular, the same $U_{\mathrm{flip}}$ unitary gate in Eq. (\ref{eq:U_flip_main_text}) can be
applied on the planes of the cube shown in Fig.~\ref{fig1}, plus all symmetric
rotations of these planes. Notice that these gates conserve the more strict
line symmetries in the plane which they are applied, but only conserve the
$U(1)$ charge on the plane for the perpendicular planes which they intersect.

We also study 3D circuit by applying the 4 sites gates
sequentially on plaquettes lying in each of the planes shown in Fig.~\ref{fig1} (plus the symmetry allowed
cubic rotations). In this case, each time step consists of 36 layers of gates
(4 layers per plane times nine planes), as opposed to the 16 layers required for
the 2D case.  

\begin{figure}
\includegraphics[scale=0.35]{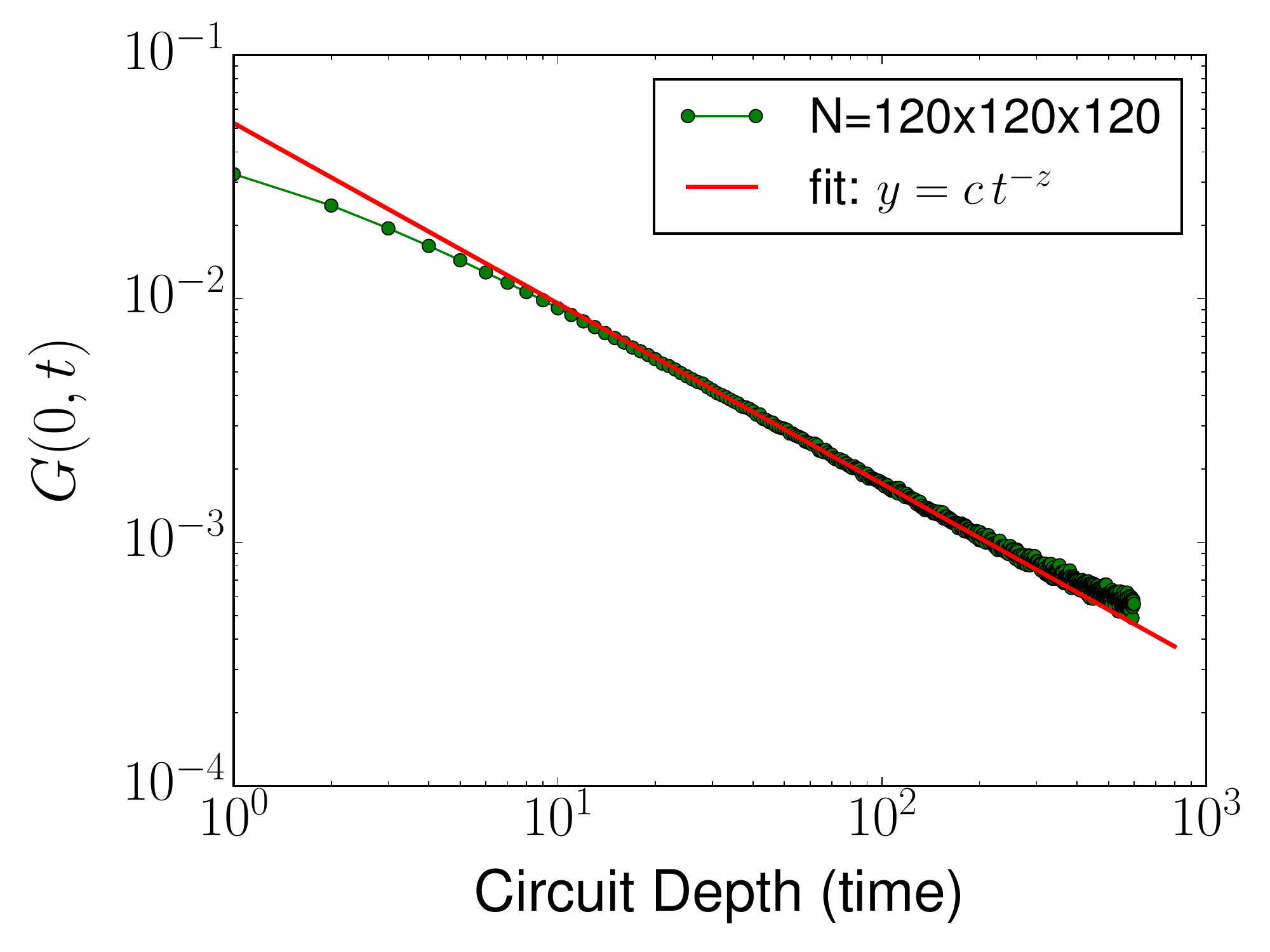}
\caption{{\bf Charge Subdiffusion with Planar Subsystem Symmetries in 3D} --  The autocorrelation function $G(0,t)$. We numerically fit this (using the intermediate times only) to the function $G(t) = c \,t^{-z}$ and find $z=0.741(6)$  and $c=.046(1)$. This is close to the analytical prediction, that $G(t) \sim t^{-0.75}$ as $N,t \rightarrow \infty$.}
\label{fig:3D_auto}
\end{figure}

We study the same quantities in the 3D model where charges are constrained to
move only along two dimensional planes. The numerics in this case find that the autocorrelation
function scales like
\begin{eqnarray} 
G(0,t)\sim \frac{1}{t^{0.741(6)}} \, \, .
\end{eqnarray}
That is, in this 3D case we find a clear violation of the diffusion law.

\begin{figure}
\includegraphics[scale=0.40]{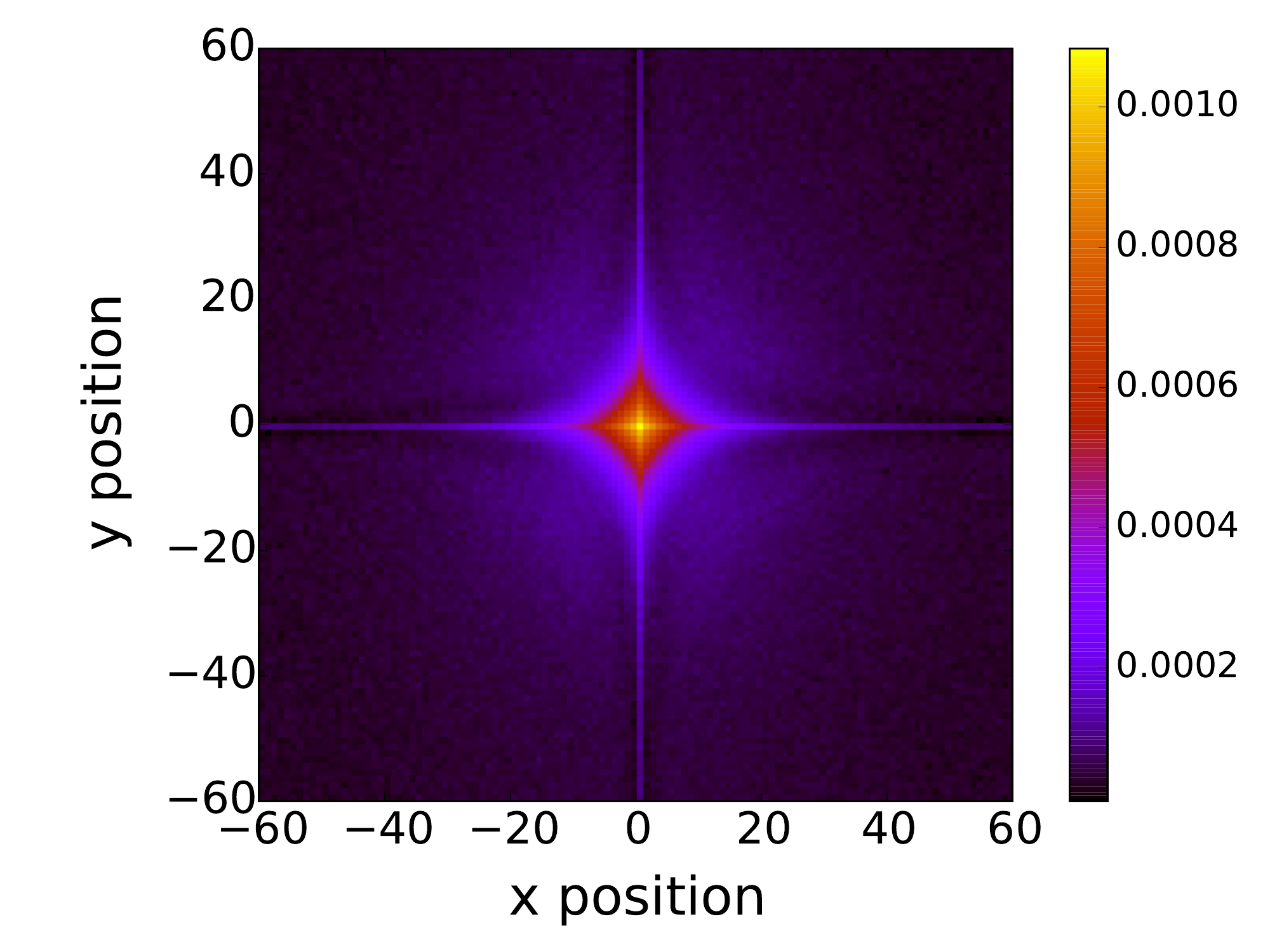}\\
\includegraphics[scale=0.345]{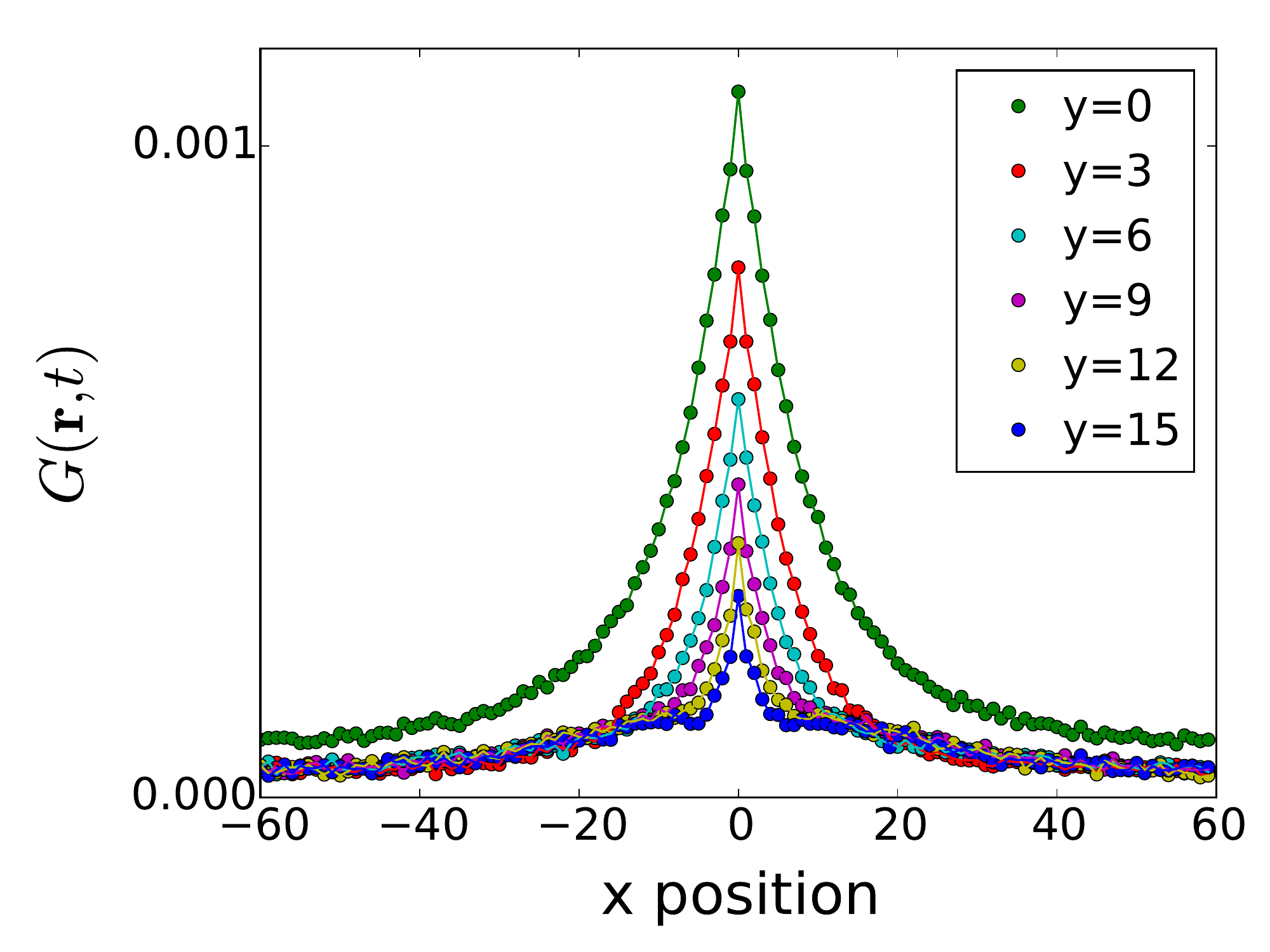}
\caption{{\bf Dynamics with Planar Subsystem Symmetries in 3D} --{\it Top:} The space time correlation function
$G(\rB, t)$ for a 2D slice of the 3D cubic lattice at
fixed time $t=200$ for a system with $N=120^3$ sites. Notice that the charge has higher density along the intersection of two planes which contain the charge. {\it Bottom:} 1D slices of the same data, plotted for different y positions. Notice that again $G(\rB,t)$ has a cusp near the origin. } \label{fig:3D_slice}
\end{figure}

We also study the spatial distribution of $G(\rB, t)$ at fixed time for a 2D slice
of the system, as shown in Fig. \ref{fig:3D_slice}. We see that again the spatial distribution is clearly not
Gaussian, and again shows a cusp near $x=0$. The charge density is
greater along the lines which are shared between pairs of planes which have nonzero
net charge.

Analytically we may understand this subddiffusive behavior by employing similar techniques as we employed in Sec. \ref{subsec:2D}. 
The conservation of planar $U(1)$ charges requires that the auto-correlation function $G_{\rB, t}$ is constrained, after summing over any $xy$, $yz$, or $xz$ plane, analogous to the constraint derived for line-like subsystem symmetries in Sec. \ref{subsec:2D}.  The simplest, discrete-time evolution of this correlation function that is consistent with these constraints, and with the symmetries of the cubic lattice is  
\begin{eqnarray}\label{eq:G_3D_main_text}
G_{\rB, t+1} &=& (1-3\lambda)G_{\rB, t} + \lambda \left[G_{\rB\pm x, t} + G_{\rB\pm y, t} + G_{\rB\pm z, t}  \right]\nonumber\\
 &-& \frac{\lambda}{4} \left[ G_{\rB\pm x\pm y, t} + G_{\rB\pm y\pm z, t} + G_{\rB\pm z \pm x, t} \right] ,
\end{eqnarray}
where $\lambda$ is, again, a free parameter of the dynamics. Taking the continuum limit of this equation in space and time yields the partial differential equation for the coarse-grained correlation function
\begin{align}
\partial_{t}G(\rB, t) = \frac{\lambda}{4}\left[\partial_{x}^{2}\partial_{y}^{2} + \partial_{y}^{2}\partial_{z}^{2} + \partial_{z}^{2}\partial_{x}^{2}\right]G(\rB, t) ,
\end{align}
which yields the result that $G(\kB, t) = \exp[-(\lambda t/4)\left[k_{x}^{2}k_{y}^{2} + k_{y}^{2}k_{z}^{2} + k_{z}^{2}k_{x}^{2}]\right]$ in momentum space.  Performing the inverse Fourier transform, and re-scaling $k \rightarrow k' = k t^{1/4}$ yields the desired result that
$G(0,t) \overset{t\rightarrow\infty}{\sim} t^{-3/4}$.

\begin{figure*}
$\begin{array}{ccc}
\includegraphics[scale=0.31]{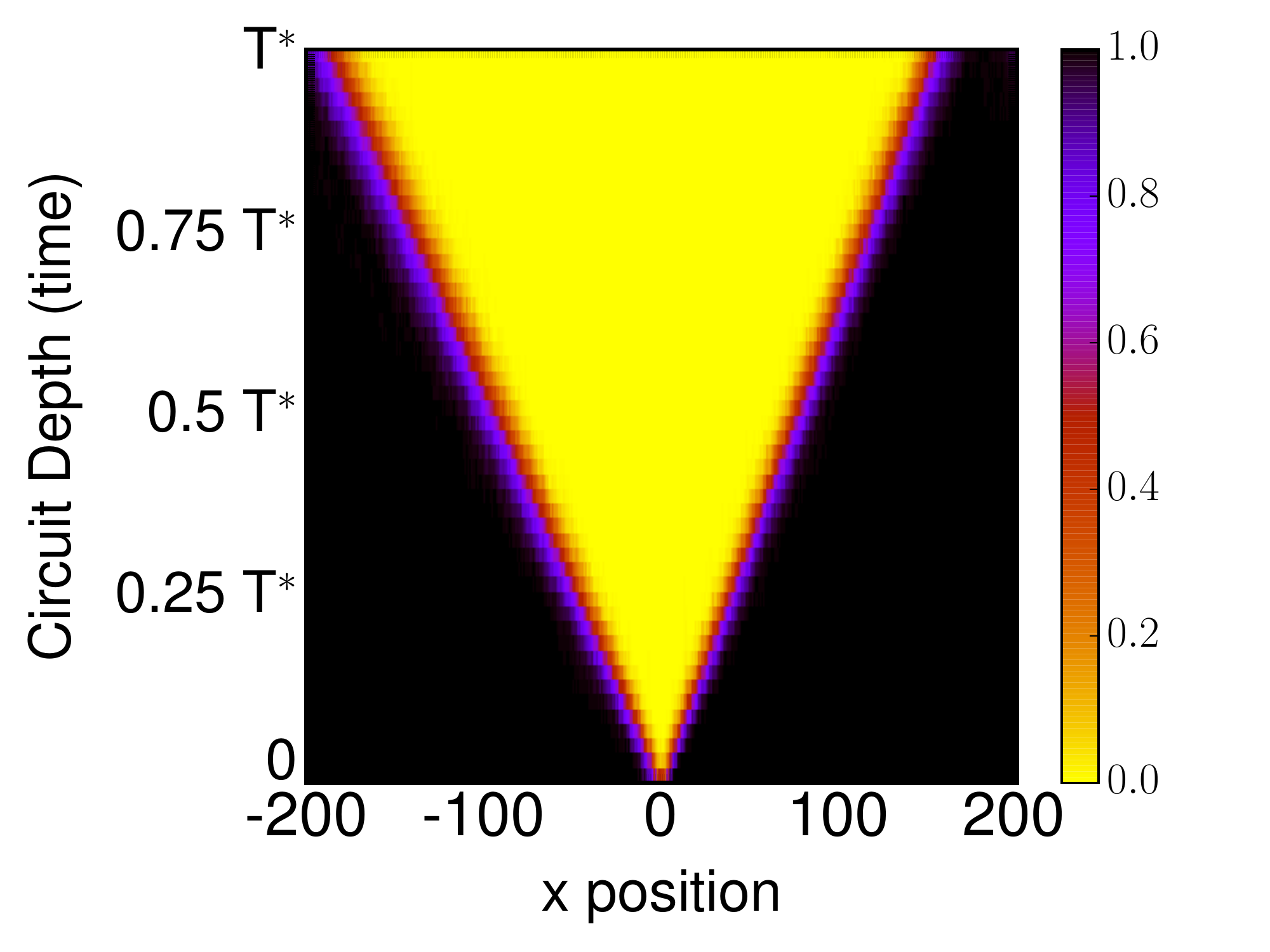} & 
\includegraphics[scale=0.3]{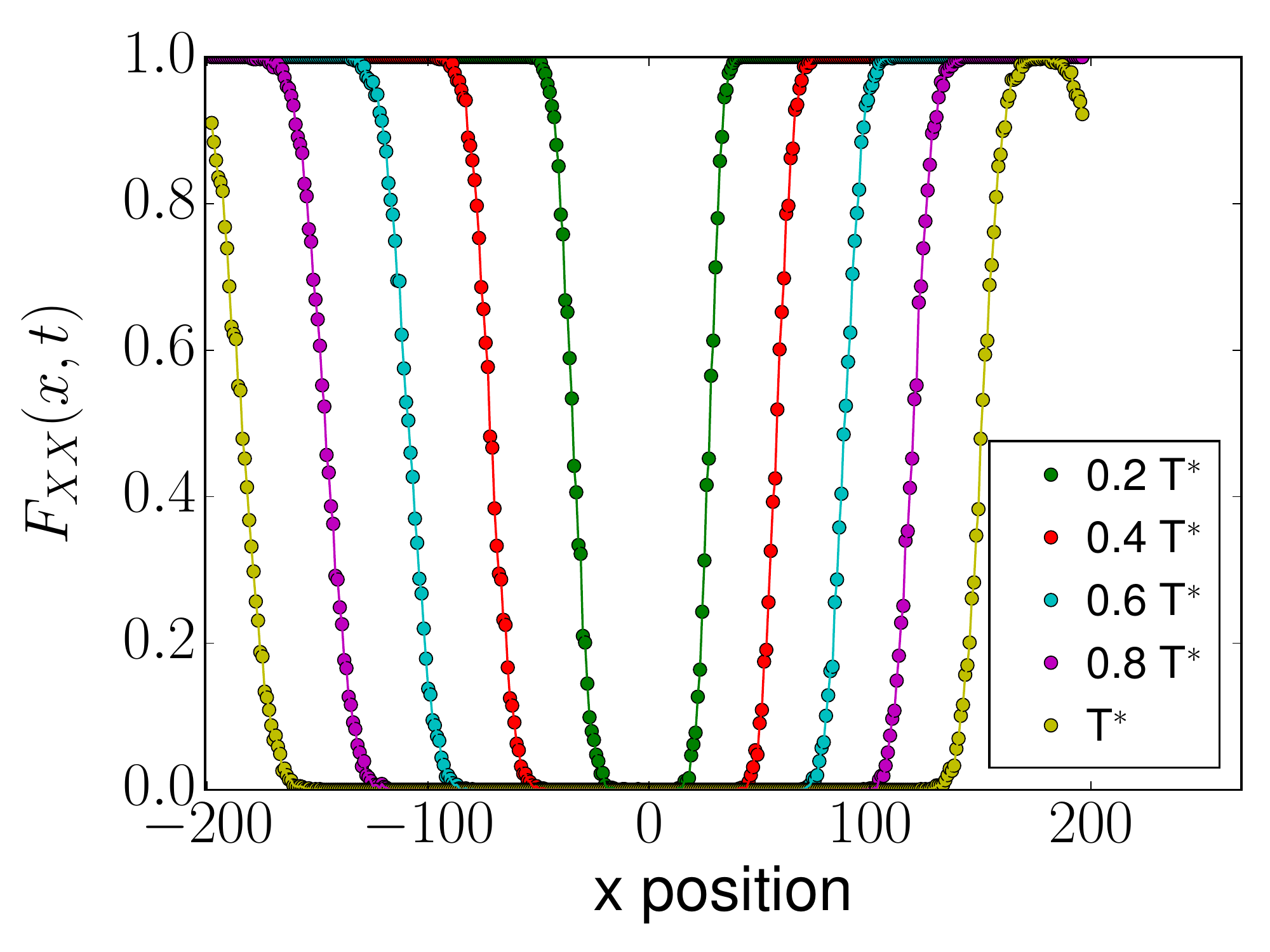} &
 \includegraphics[scale=0.29]{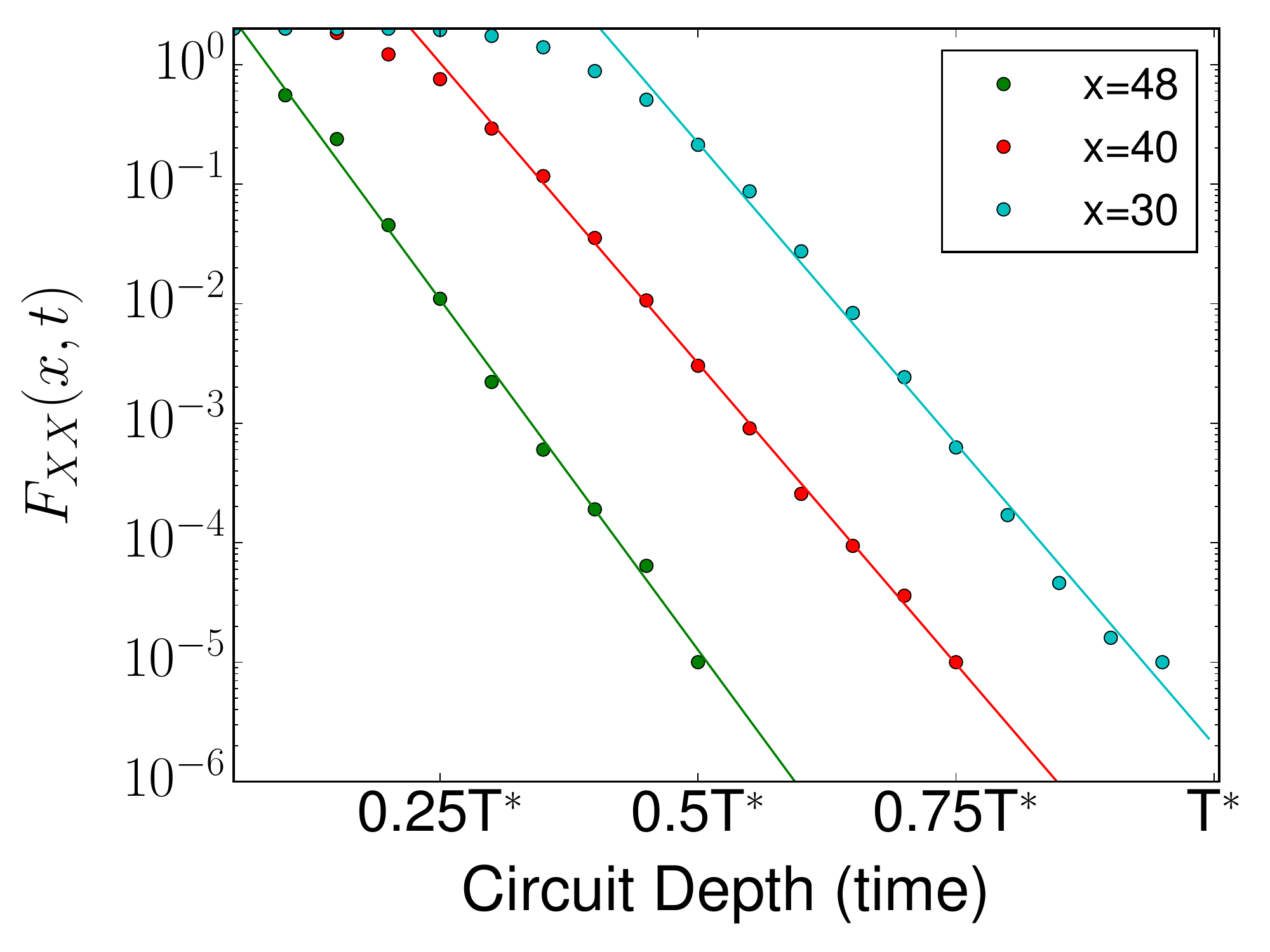}\\
 \text{(a)} & \text{(b)} & \text{(c)}
  \end{array}$
\caption{{\bf XX OTOC for the 2D Automaton Circuit} -- We measure the XX OTOC along the line $\rB = (x,0)$ for the 2D automaton circuit.  In (a) and (b), we observe a clear light cone, for a system with $N=396^2$ sites, whereby
$F(\rB,t)\approx 1$ for $|\rB|>v_Bt$ and $F(\rB,t)\approx 0$ for $|\rB|<v_B t$, as well as a broadening of the ballistically-propagating front of the operator. In (c), we observe that at a fixed spatial position, the OTOC decays  exponentially to zero.  Shown are exponential curves $\sim \exp[-c t]$ where $c\sim 1.53 - 1.16$ depending on the position x. For this plot we simulated a smaller system with $N=96^2$ sites in order to resolve this exponential decay with greater accuracy.  In all plots, $T^* = {L}/({2v_B})$ is the time for the lightcone to spread across the entire lattice. }
\label{OTOC1}
\end{figure*}

As before, a more careful analysis of the difference equation (\ref{eq:G_3D_main_text}), where we only take the continuum limit in time, while Fourier transforming in space, yields
\begin{align}\label{eq:G_3D_k_space}
&\frac{d}{dt}G(\kB,t) = - f(\kB) G(\kB,t)\\
f(\kB) &= 4\lambda \sum_{i\ne j\in\{x,y,z\}} \sin^2\left(\frac{k_{i}}{2}\right) \sin^2\left(\frac{k_{j}}{2}\right).
\end{align}
While we are unable to analytically perform the inverse Fourier transform to obtain the exact autocorrelation function from these expressions, we observe that $f(\kB)$ vanishes along the three lines $\kB = (k_{x},0,0)$, $(0,k_{y},0)$, $(0,0,k_{z})$ that meet at the origin.   As a result, the spatial continuum limit considered previously --  which corresponds to an expansion of  (\ref{eq:G_3D_k_space}) near $\kB \rightarrow 0$, where  $f(\kB)$ vanishes  most rapidly -- is justified in understanding the long-time asymptotics of $G(0, t)$. 

\section{Operator Spreading in Automaton Circuits} 
\label{chaos}

We now numerically study the behavior of the out-of-time-ordered correlation function (OTOC) -- defined in Eq. (\ref{eq:OTOC_main_text}) as  $F(\rB, t) = ({1}/{D})\text{Tr}\left[\mO_{\rB}(t)\mO'_{0}\mO_{\rB}(t)\mO'_{0}\right]$ -- in order to quantify how local operators evolve under Heisenberg evolution by the automaton quantum circuits that we have considered.  The primary purpose of our study of the OTOC is to argue that the ``non-conserved part" of Heisenberg evolution of a local operator under automaton unitary dynamics -- defined as the portion of the evolution of an operator that has no overlap with the conserved charges in the evolution --  is quite generic, and resembles the evolution under the most general, symmetry-preserving unitary gates.   

We study the following out-of-time-ordered correlation functions
\begin{align}
F_{ZX}(\rB, t) &\equiv \frac{1}{D}\Tr[Z_{\rB}(t)X_{0} Z_{\rB}(t) X_{0}] \label{eq:ZX_OTOC} \\
F_{XX}(\rB, t) &\equiv \frac{1}{D}\Tr[X_{\rB}(t)X_{0} X_{\rB}(t) X_{0}], \label{eq:XX_OTOC}
\end{align}
which we refer to as the ``ZX" and ``XX" OTOC's for the remainder of this section.

As detailed in Sec. \ref{sec:q_chaos_sec},  the ballistic growth of a non-conserved operator in a chaotic quantum system, and the equilibration of the local structure of these operators implies ($i$) that the OTOC has a ballistically-propagating front, and ($ii$) that $F(\rB, t) \rightarrow 0$ as $r/(v_{B}t)\rightarrow0$.  The saturation of the OTOC also coincides with the development of volume-law ``operator entanglement" \cite{Chaos_Qi}.  Finally, we observe that ($iii$) the front of the OTOC is believed to broaden as $\sim t^{\alpha}$, where $\alpha$ is a dimension-dependent exponent; in one spatial dimension, the ends of an operator perform a biased random walk, leading to $\alpha = 1/2$, while in higher dimensions, these exponents are related to the probability distributions for classical stochastic growth processes \cite{PhysRevX.8.021014}.  We will verify these three features, which are characteristic of operator spreading in chaotic systems, in the out-of-time-ordered correlations of the automaton circuits considered previously.    

In addition, we expect the ZX OTOC to have ``tails" connecting the ballistically propagating front of the OTOC, to its value at position $\rB$, due to the fact that the slow-moving conserved operators appearing in the Heisenberg evolution of $Z_{\rB}$ can ``emit" non-conserved operators, which then propagate ballistically \cite{KhemaniVishHuse}.   In contrast, no such tails are expected for the XX OTOC, which should saturate (exponentially) rapidly to its asymptotic value at times $t>r/v_{B}$.  As a result, we consider a scaling form for the XX OTOC,  
\begin{align}\label{eq:scaling_collapse_OTOC}
F_{XX}(\rB, t) = f\left( \frac{r - v_{B}(\Omega)t}{t^{\alpha(\Omega)}}\right),
\end{align}
which is motivated by similar scaling forms for the OTOC in a random unitary circuit \cite{PhysRevX.8.021014}. Here, $v_{B}(\Omega)$ and $\alpha(\Omega)$ are direction-dependent butterfly velocities and broadening exponents, respectively, and $\Omega$ denotes a $d$-dimensional angle.

 We measure these out-of-time-ordered quantities using classical Monte Carlo techniques. For example, we may write the ZX OTOC as
\begin{align}
F_{ZX}(\rB, t) &= \frac{1}{D}\sum_n \bra{n} U^\dagger Z_{\rB} U X_{0} U^\dagger Z_{\rB}
U X_{0} \ket{n} \nonumber\\
&= \frac{1}{D}\sum_{n} \bra{n(t)}Z_{\rB}\ket{n(t)} \bra{n'(t)} Z_{\rB} \ket{n'(t)} ,
\end{align}
where $\ket{n'} = X_{0} \ket{n}$. 
Therefore, measuring the ZX OTOC is as simple as keeping track of the
classical evolution of the two states $\ket{n}$, $\ket{n'}$.  The XX OTOC may be measured in a similar manner,  by evolving the two states $\ket{n}$
and $\ket{m}$, applying a flip operator at site $\rB$ to each state, and 
evolving the resulting states backwards in time. The average overlap between the
resulting states gives the XX OTOC.  In this case, every space-time measurement at point $(\rB,t)$ requires an independent simulation, making the XX OTOC measurements somewhat slower to simulate in practice.

\subsection{Numerical Study of the OTOC}
We now present the results of our measurements of the out-of-time-ordered
correlator. 


\begin{figure}
\includegraphics[scale=0.29]{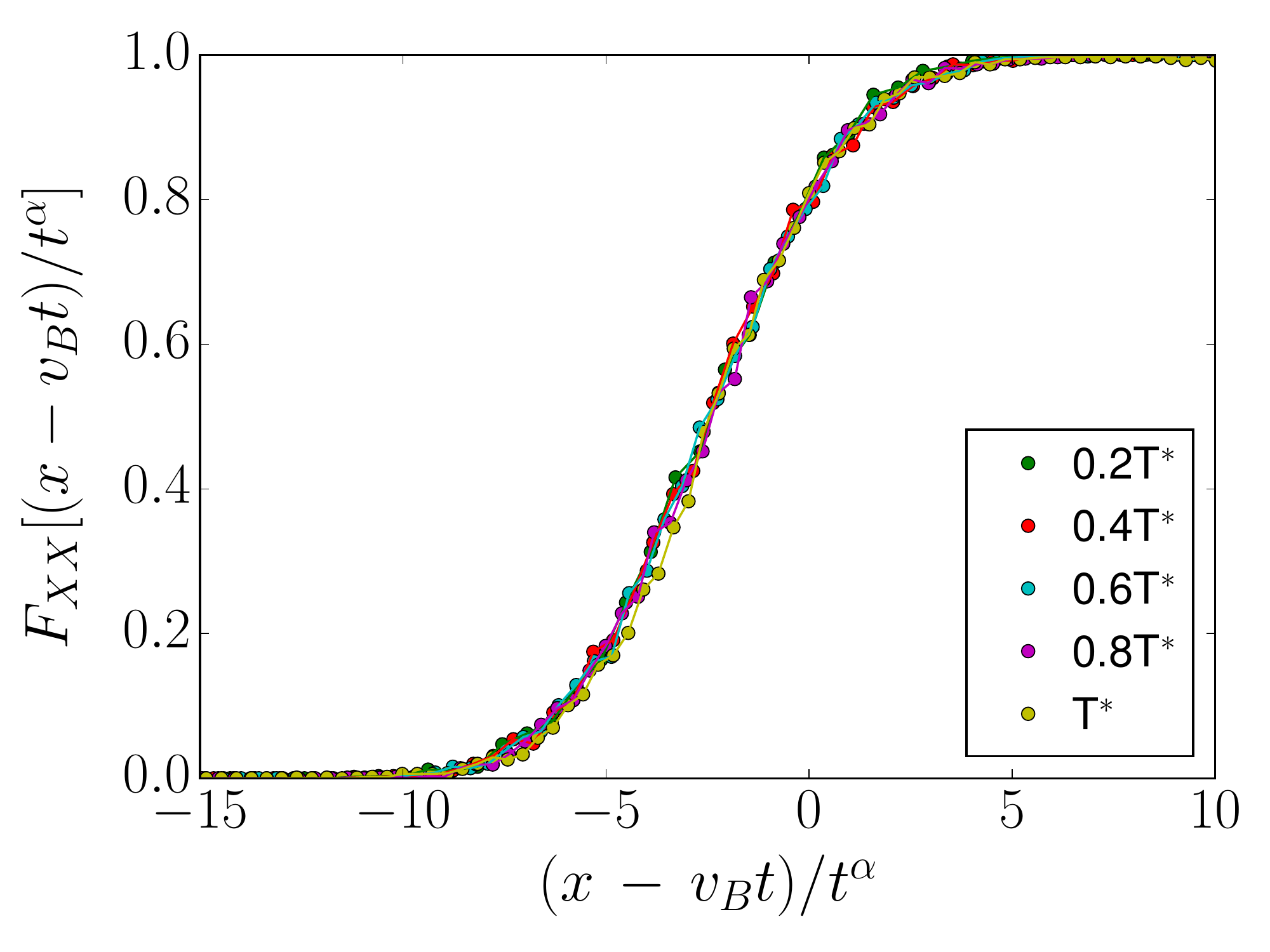} 
\includegraphics[scale=0.29]{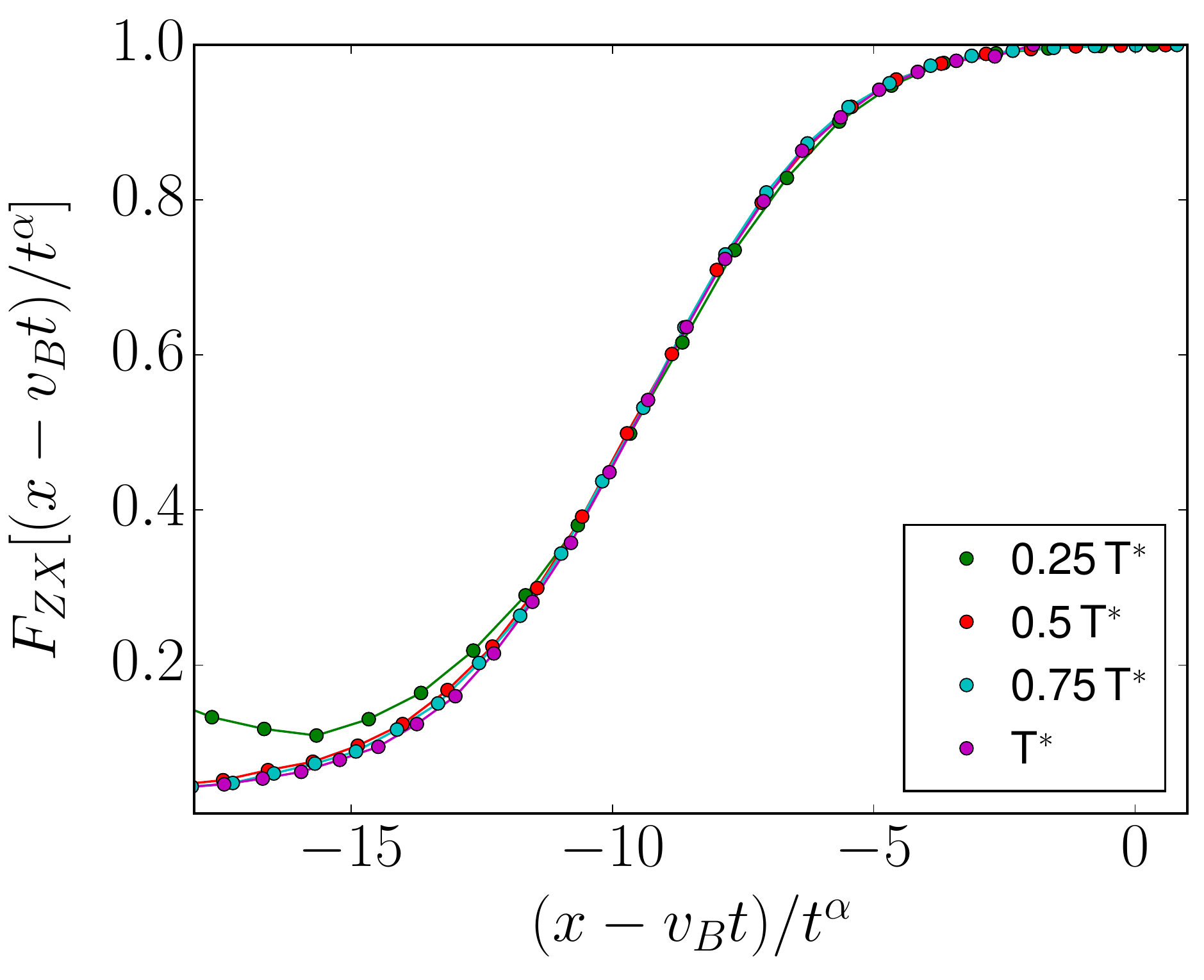} \\
\caption{{\bf Scaling Collapse of the OTOC for the 2D and 3D Automaton Circuits} --  {\it Top: } Shown is a scaling collapse of $F_{XX}(x,t)$ in 2D to the functional form in Eq. (\ref{eq:scaling_collapse_OTOC}), from which we observe that the operator front broadens as $\sim t^{\alpha}$ with $\alpha = 0.308(18)$. {\it Bottom: } The scaling collapse of the front of $F_{ZX}(x,t)$ in 3D with $\alpha=0.220(5)$. These should be compared with the theoretically predicted values of $\alpha = 1/3$ and $\alpha=0.24$ for the 2D and 3D cases respectively. Again, we take $T^*={L}/({2v_B})$. Both OTOC's are calculated along the $x$ axis.}
\label{OTOC2} 
\end{figure}

In Fig.~\ref{OTOC1}, we see the growth of the XX OTOC along the line $\rB = (x,0)$, for our 2D automaton circuit. We observe a ballistically
spreading light cone, and that the OTOC is very nearly zero inside the light cone and one
outside.  From similar measurements of the XX OTOC along other directions, we observe a nearly isotropic butterfly velocity.  We also see in Fig.~\ref{OTOC1}c, that for a fixed position, the OTOC appears to decay exponentially in time, within the times that we are able to perform numerical simulations.  This is consistent with the expectation that the OTOC should decay to zero at sufficiently long times.   Finally, shown in Fig.~\ref{OTOC2} is a   scaling collapse of the XX OTOC, to the functional form in Eq.~\ref{eq:scaling_collapse_OTOC}; the optimal scaling collapse is to the butterfly velocity $v_{B}$ and front broadening exponent $\alpha$, which are given by 
\begin{align}
\frac{v_{B}}{v_{\mathrm{max}}} \approx 0.58 \hspace{.5in} \alpha = 0.308(18).
\end{align}
Here, $v_{\mathrm{max}}$ is the lightcone velocity along the line $\rB = (x,0)$ for our automaton circuit.   This should be compared with $\alpha = 1/3$, which is the observed broadening for a two-dimensional, random unitary circuit, for which the growth of the OTOC is related to the stochastic growth of a two-dimensional cluster.  The fluctuations in the growing edge of the cluster are believed to be the same as the fluctuations in the height of a stochastically growing, one-dimensional interface, which is known to grow in time as $t^{1/3}$ \cite{PhysRevX.8.021014}. 

\begin{figure}[t]
\includegraphics[scale=0.30]{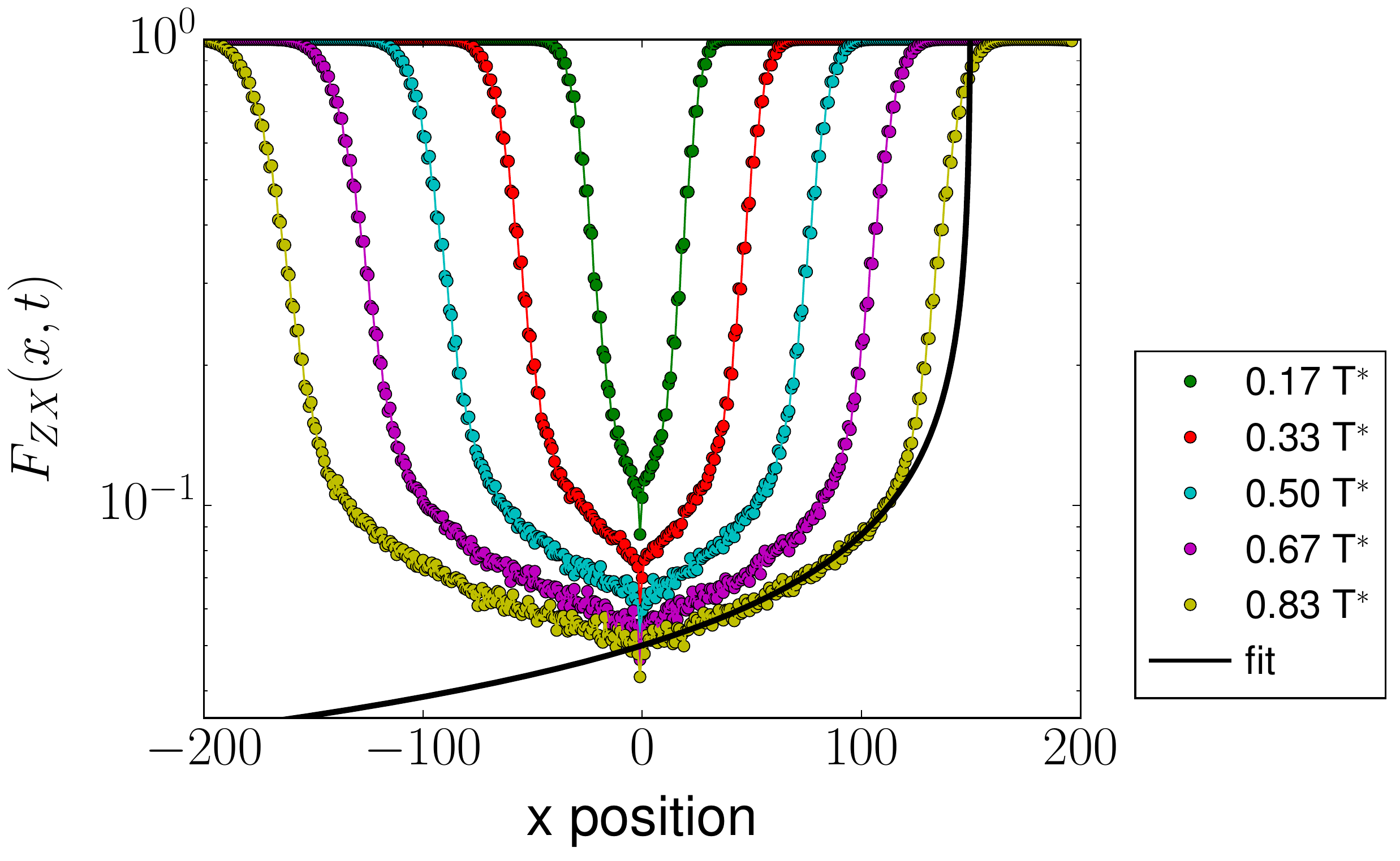} \\
\includegraphics[scale=0.30]{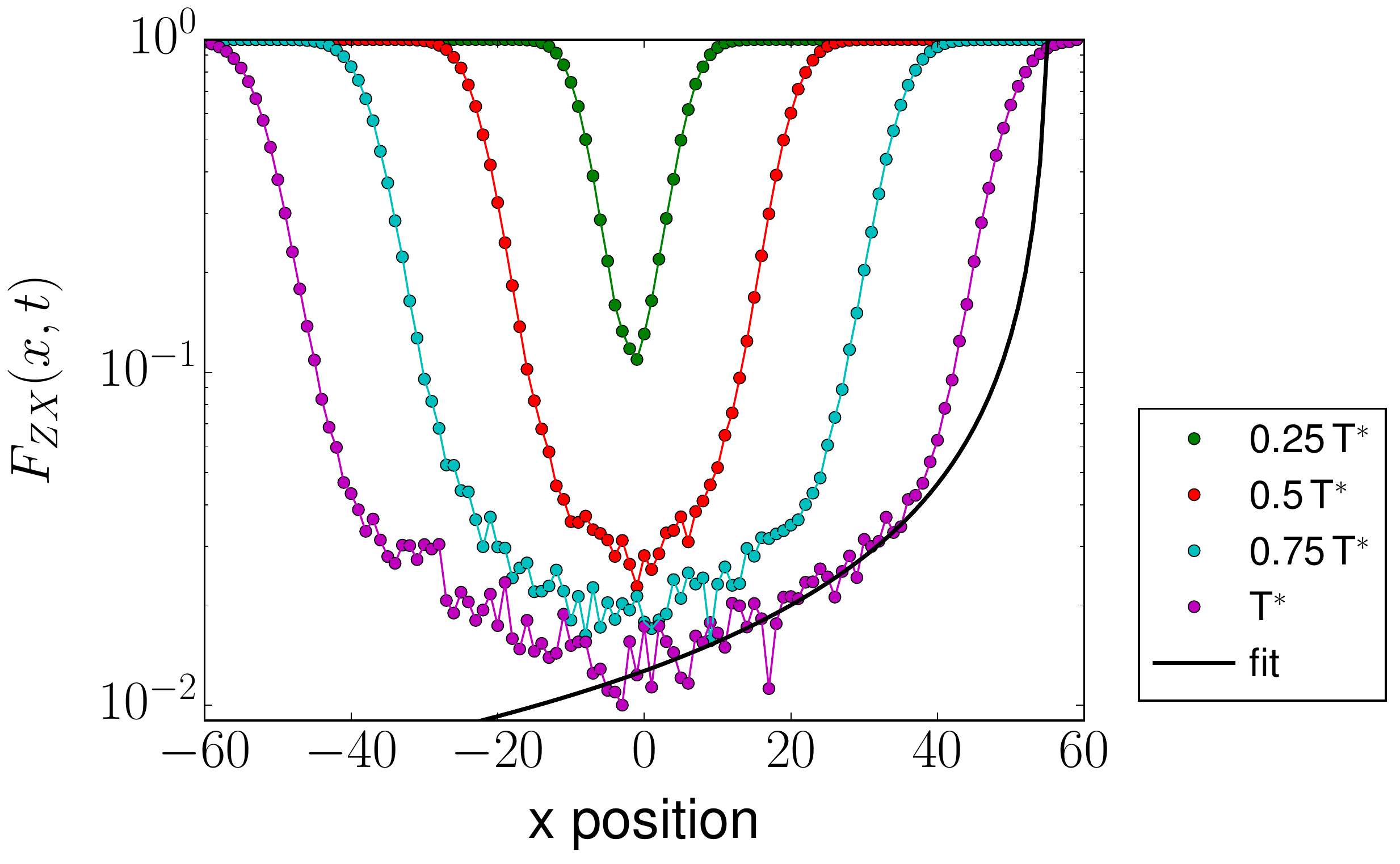}
\caption{{\bf The ZX OTOC } -- The OTOC $F_{ZX}(\rB, t)$ is shown along the $x$ axis, for both the 2D (top) and 3D (bottom) automaton circuits considered previously. Here we can see
a clear tail behind the leading OTOC wave front. In the 2D circuit,
this tail appears to have the same functional form as a sub-leading tail for a 1D circuit
with a global $U(1)$ symmetry. Fitting the tail at the latest times to $F_{ZX} \sim (v_B t-|\rB|)^{-\beta}$ gives $\beta = 0.50(1)$. For the 3D
circuit, the tail of the OTOC (when $v_{B}t\gg |\rB|$) takes the form $F_{ZX} \sim (v_B t - |\rB|)^{-\beta}$ with $\beta =1.02(7)$. Again, we take $T^*={L}/({2v_B})$.}
\label{OTOC3}
\end{figure}

In Fig.~\ref{OTOC3}, we look at the ZX OTOC, and we observe a ``tail" behind the 
ballistically-propagating front of the OTOC, due to the fact that the slow-moving operators that have an overlap on the subsystem symmetry charges continue to ``emit" non-conserved operators that propagate ballistically. We fit the tail of the OTOC data in two and three dimensions to the functional form 
\begin{align}\label{eq:tail}
F_{XX}(\rB, t) \overset{v_{B}t\gg |\rB|}{\sim} \frac{1}{\left(v_{B}t - |\rB|\right)^{\beta}}
\end{align}
and find that $\beta = 0.50(1)$ for our 2D circuit with line-like subsystem symmetries, while $\beta = 1.02(7)$ for the 3D circuit with planar symmetries.  Trendlines with this power-law behavior are shown in Fig. \ref{OTOC3}.   

For the 3D case, we are unable to calculate the XX OTOC to long enough times that we are able to perform a satisfactory scaling collapse to quantify the broadening of the front of a non-conserved operator.  Instead, we restrict our attention to the ZX OTOC along the line $\rB = (x,0,0)$.  We obtain a convincing scaling collapse of the ballistically propagating front of this OTOC -- with a front broadening exponent $\alpha = 0.220(5)$ -- in Fig. \ref{OTOC2}b; this is very close to the predicted value of the broadening exponent obtained from studies of operator spreading in Haar-random  unitary circuits \cite{PhysRevX.8.021014}, which predict $\alpha=0.24$. We also observe that the tail follows a power law form as in Eq.~(\ref{eq:tail}), except with exponent $\beta = 1.02(7)$, as shown in Fig. \ref{OTOC3}.

\subsection{Recurrence times}

\begin{figure}[t]
\includegraphics[scale=0.30]{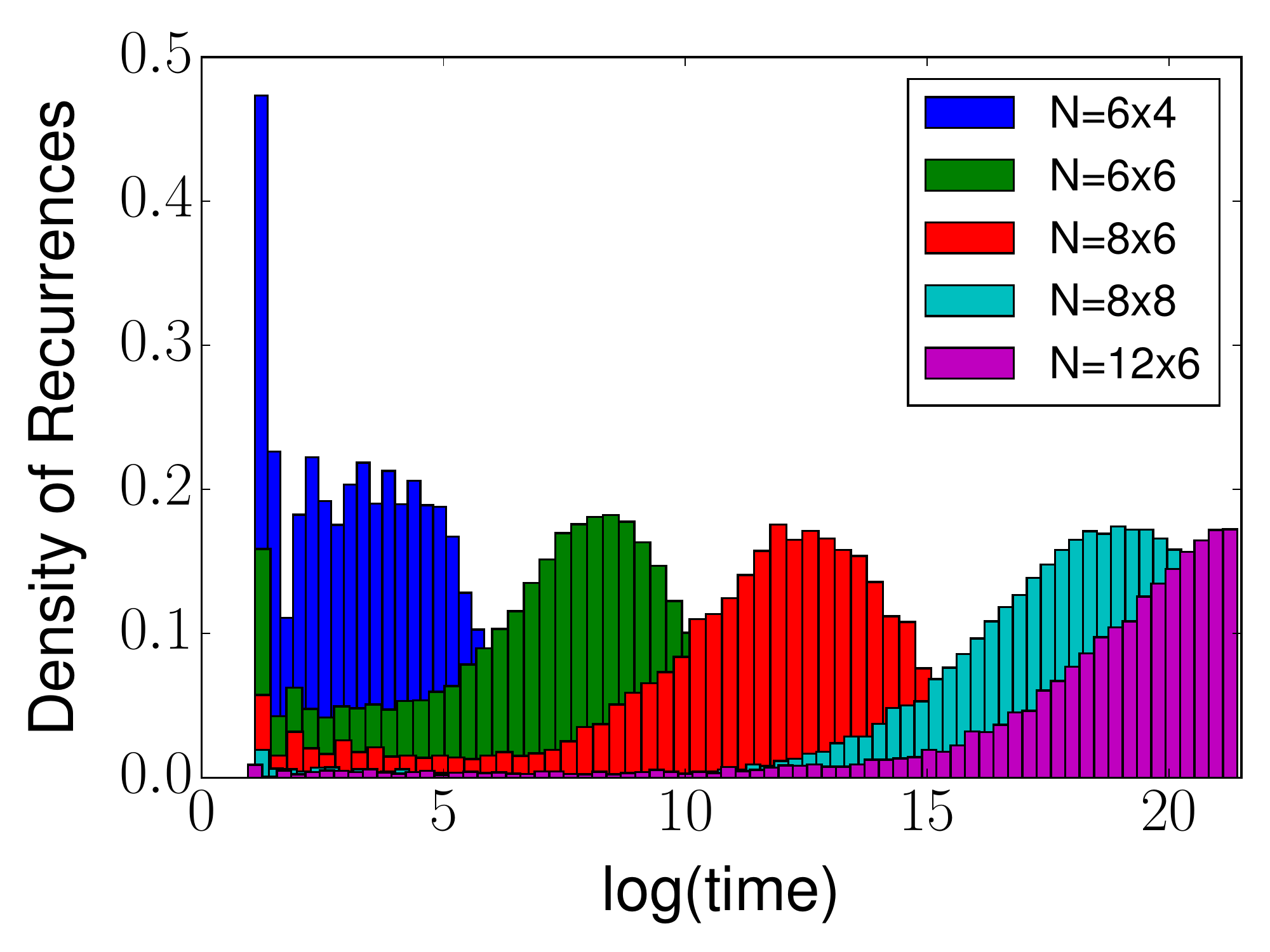}\\
\includegraphics[scale=0.30]{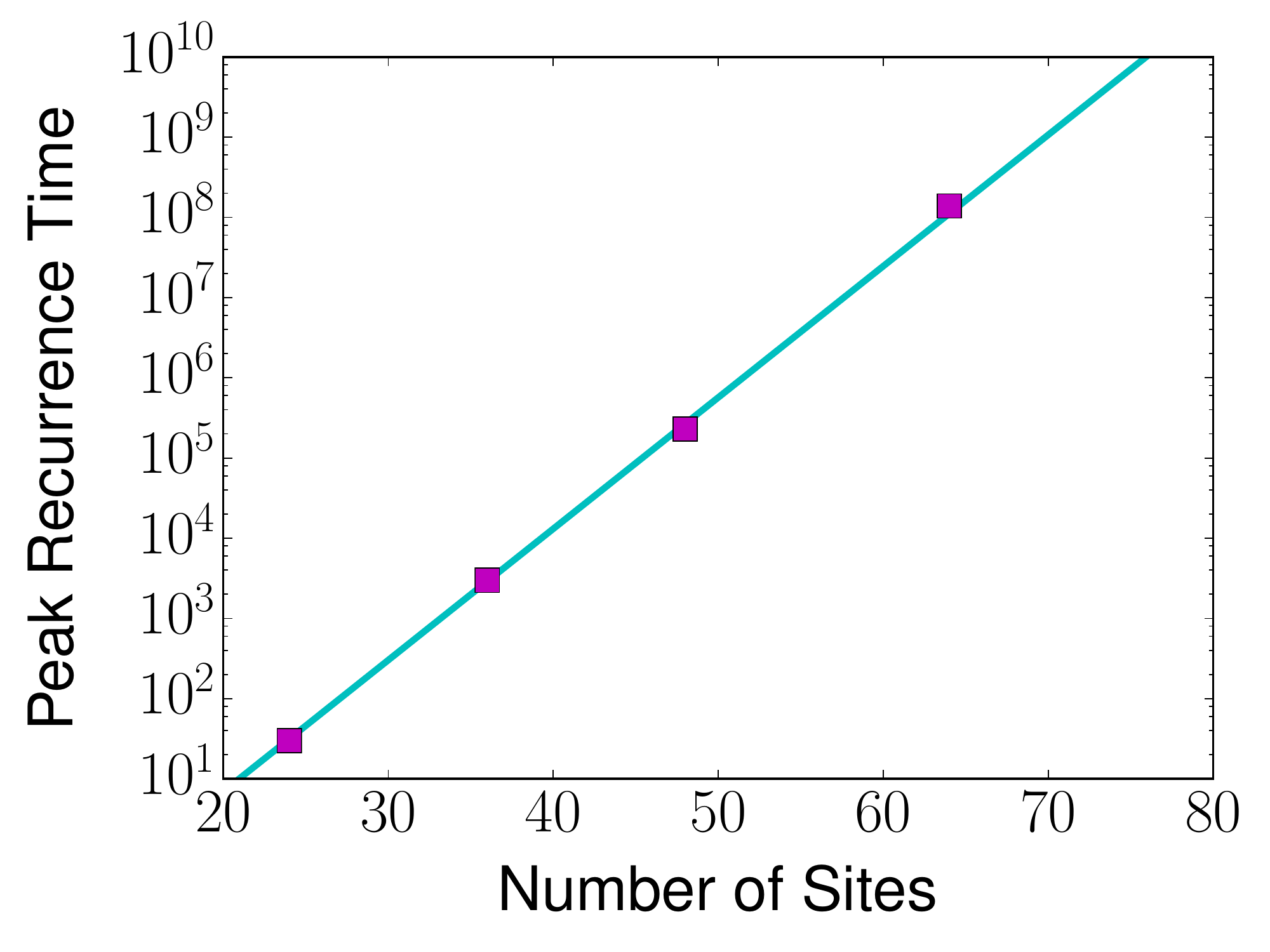}
\caption{{\bf Recurrence Times} -- The number of time steps that the 2D circuit is run before an initial
product state $\ket{n}$ returns to its initial value $U^t\ket{n} = \ket{n}$. The
average number of time steps needed to see such a recurrence grows exponentially
with the volume of the system. }
\label{recurrence}
\end{figure}

Finally, we study the recurrence time for product states in the Pauli $Z$ basis, evolving under the automaton dynamics; these product states only evolve to other
product states in the same basis. The recurrence time $t_{\mathrm{prod}}(n)$ is then the
circuit depth for which the state returns to its initial value, i.e. 
$U^{t_{\mathrm{prod}}(n)}\ket{n} = \ket{n}$.  In Fig.~\ref{recurrence}, we show the
distribution of these recurrence times, taken over the full set of states $\ket{n}$ in the
computational basis, for different system sizes. 
From the peaks of these distributions, we obtain the typical recurrence time as a function of 
system size, as shown in Fig.~\ref{recurrence}b). We see that $t_{\mathrm{prod}}$ grows linearly in the Hilbert space dimension $D$ of the system, as predicted for a random automaton dynamics. This is considerably slower than for fully chaotic systems, where the recurrence time would be doubly exponential in the number of sites, but considerably faster than  translation invariant, Floquet Clifford circuits, where one can prove that $t_{rec} = \mathcal{O}(\log D)$ \cite{Keyserlingk2}. 

\section{Conclusions}
\label{conclude}
We have discussed a new class of circuits in two and three dimensions, which produce a quantum dynamics that is { classically simulable}. In particular, the evolution of certain correlation functions under the action of these circuits is classically simulable even though the circuits generate operator entanglement, since these dynamics avoid generating {state} entanglement when acting on product states in the computational basis. We have introduced a set of circuits that respect {\it subsystem symmetries}, conserving a $U(1)$ charge along either { lines} in two space dimensions, or along { planes} in three space dimensions. These constraints lead to dynamics that lies in a new universality class, with charge spreading {\it subdiffusively} both with respect to the physical dimension and with respect to the dimension of the submanifold in which the symmetries act. In particular, for a two dimensional system with line like symmetries, charge spreads as $\log(t)/\sqrt{t}$, whereas for a three dimensional system with planar symmetries, charge spreads as $1/t^{3/4}$. This behavior is captured by a finite difference equation, the continuum limit of which yields an `anomalous subdiffusion' equation in which the Laplace operator is replaced by something akin to a square of the Laplace operator.

While the dynamics we have considered are structured (and have to be, in order to be classically simulable), we have nevertheless shown that they do exhibit several features commonly associated with chaos. In particular, OTOC's involving non-conserved operators exhibit a ballistic lightcone with a front that broadens as a dimension-dependent power-law.  Furthermore these OTOC's appear to saturate to zero at long times, which suggests that the bulk structure of a non-conserved operator equilibrates in a manner that is consistent iwth quantum chaos.  OTOCs between a conserved and non-conserved operator exhibit in addition a `universal tail' behind the propagating front, as has also been observed in a random circuit with a global $U(1)$ symmetry\cite{KhemaniVishHuse}. As such, we conjecture that our results may hold in more generic quantum dynamics with the same symmetries. 

This work opens up a new direction for the exploration of quantum dynamics, rendering accessible questions regarding chaos and operator spreading in {\it higher} dimensions in a class of circuits that are much less structured and more generic than any previously known classically-simulable higher dimensional circuit of which we are aware. It also opens the door to a new class of dynamical phenomena produced by subsystem conservation laws. 

A number of extensions of this work immediately present themselves for consideration. One possibility is to move away from the square/cubic lattices considered herein, and to consider instead subsystem symmetries on more general crystal lattices. Symmetries involving fractal subsystems, as in Type II fracton models \cite{fracton2, haah} are important to consider. Either of these extensions holds the promise of qualitatively new universality classes for dynamical behavior. Extensions of this work to {\it Hamiltonian} systems (rather than quantum circuits) are of interest. We expect analogous behavior in the Hamiltonian setting.  Other potential extensions include considering subsystem symmetry groups other than $U(1)$. A final possibility is to move away from gates with strictly local support to e.g. quasilocal gates with exponential tails in real space. We leave such questions for future work. 

\begin{acknowledgments} 
This material is based upon work supported by the Air Force Office of Scientific Research under award number FA9550-17-1-0183 (JI and RMN). SV is supported by the Harvard Society of Fellows. SV and RMN also acknowledge the hospitality of the KITP, where this work was initiated, during a visit to the program {\it Dynamics of Quantum Information.} The KITP is supported in part by the National Science Foundation under Grant No. NSF PHY-1748958.

\end{acknowledgments}

\appendix
\section{Operator Spreading under Automaton Dynamics}\label{appendix:op_spreading}
It is convenient to study the growth of the complexity of an operator evolving under the automaton dynamics as follows.  We consider $N$ spins with Hilbert space dimension $D = 2^{N}$. Let $n\in\{0,1,\ldots,D-1\}$ be an integer, and let $\boldsymbol{n} = (n_{1},n_{2},\ldots,n_{N})$ be an $N$-component vector that corresponds to the binary representation of integer $n$.  We further define $\mS_{n}$ to be a product of $Z$ operators that corresponds to the binary representation $n$.  Specifically, 
\begin{align}
\mS_{n} \equiv \prod_{r=1}^{N} \left(Z_{r}\right)^{n_{r}}
\end{align}
For example, $\mS_{0}$ is simply the identity operator, while $\mS_{7} = Z_{1}Z_{2}Z_{3}$ since seven is represented as $111000\cdots$ in binary.  We note that $\frac{1}{D}\Tr(\mS_{n}\mS_{m}) = \delta_{nm}$ where $D$ is the total Hilbert space dimension. 

Under the automaton dynamics, these operators evolve as 
\begin{align}\label{eq:automaton_dynamics}
\mS_{n}(t) = U(t)\mS_{n}U(t)^{\dagger} = \sum_{m=0}^{D-1}a_{nm}(t) \,\mS_{m}.
\end{align}
In contrast, for a more general unitary dynamics, each of the operators $\{\mS_{n}\}$ will evolve into sums of  arbitrary products of Pauli operators in the system.  

It is now convenient to define the following wavefunction in a spin-(1/2) system with the same total Hilbert space dimension
\begin{align}\label{eq:n}
\ket{n(t)} \equiv \sum_{m=0}^{D-1} a_{nm}(t) \ket{m}
\end{align}
where $\ket{m}$ is the representation of integer $m$ as the state of a spin-(1/2) system in the Pauli-$Z$ basis, with an up (down) spin appearing whenever a 0 (1) occurs in the binary representation of $m$; as a concrete example, the state 
\begin{align}\label{eq:1_binary_state}
\ket{1} \equiv \ket{\downarrow\uparrow\uparrow\cdots}
\end{align}
To summarize, Eq.  (\ref{eq:automaton_dynamics}) describes the Heisenberg evolution of operators under the automaton dynamics, while Eq.  (\ref{eq:n}) describes a fictitious dynamics of a many-body wavefunction in the same Hilbert space.  While the entanglement of a wavefunction in the $Z$ basis does not grow under the automaton dynamics, the states of the fictitious spin system do become highly entangled as they evolve, reflecting the fact that typical operators become quite entangled under the automaton evolution. 

We now identify a few key points about the fictitious dynamics of the state $\ket{n(t)}$. 
\begin{enumerate}[{\bf (I)}]
\item The evolution of this wavefunction is {unitary}, so that
\begin{align}\label{eq:n_evolution_unitary}
\ket{n(t)} = W(t) \ket{n}.
\end{align}
where $W(t)W(t)^{\dagger} = W(t)^{\dagger}W(t) = 1$.

\item  $W(t)$ is exactly
\begin{align}
W(t) = H^{\otimes N} U(t) H^{\otimes N}
\end{align}
where $H$ is the single-qubit Hadamard gate that acts on Pauli matrices as $HXH = Z$, $HZH = X$.  That is $W(t)$ is the \emph{same} automaton unitary operator acting in a rotated basis.
\end{enumerate}

\underline{ Proof of {\bf (I)}:} Since $\langle{m|W(t)|n}\rangle = a_{nm}(t)$, we observe that 
\begin{align}
\bra{m}W(t)^{\dagger} W(t)\ket{n} &= \sum_{k} a_{mk}(t)a_{nk}(t) \nonumber\\
&= \frac{1}{D}\Tr[\mS_{m}(t)\mS_{n}(t)] = \delta_{mn}\\
\bra{m}W(t) W(t)^{\dagger}\ket{n} &= \sum_{k} a_{km}(t)a_{kn}(t) \nonumber\\
&= \frac{1}{D}\Tr[\mS_{m}(-t)\mS_{n}(-t)] = \delta_{mn}
\end{align}
so that $W(t)$ is indeed unitary.\\

\underline{ Proof of {\bf (II)}:} The automaton unitary $U(t)$ acts as a permutation $\pi\in S_{D}$ on the $D$-dimensional set of product states in the Pauli $Z$ basis.  
We define the projection operator
\begin{align}\label{eq:projector_appendix}
P_{n}\equiv |n\rangle \langle n|
\end{align}
The automaton unitary acts as
\begin{align}
U(t)^{\dagger}P_{n}U(t) = P_{\pi(n)}
\end{align}
This relation places a restriction on the coefficients $a_{mn}(t)$.  We first rewrite the projection operator as  
\begin{align}
P_{n} &= \prod_{r = 1}^{N}\left[\frac{1 + (-1)^{{n}_{r}}Z_{r}}{2}\right]
 =\frac{1}{D}\sum_{m=0}^{D-1}(-1)^{\boldsymbol{n}\cdot\boldsymbol{m}}\mS_{m}
\end{align}
where $\boldsymbol{n} = (n_{1}, \ldots, n_{N})$ is the $N$-component binary vector that corresponds to the binary representation of $n$, and where $\boldsymbol{n}\cdot\boldsymbol{m}$ is the dot product of the two vectors.  The Heisenberg evolution of this projector may be written as
\begin{align}
U(t)^{\dagger}P_{n}U(t) &=\frac{1}{D}\sum_{m=0}^{D-1}(-1)^{\boldsymbol{n}\cdot\boldsymbol{m}} U(t)^{\dagger}\mS_{m}U(t)\nonumber\\
&= \frac{1}{D}\sum_{m,k=0}^{D-1}(-1)^{\boldsymbol{n}\cdot\boldsymbol{m}}a_{mk}(t) \,\mS_{k} = P_{\pi(n)}\nonumber
\end{align}
so that
\begin{align}
\sum_{m = 0}^{D-1}(-1)^{\boldsymbol{n}\cdot\boldsymbol{m}}a_{mk}(t) = (-1)^{\boldsymbol{\pi}(n)\cdot\boldsymbol{k}}
\end{align} 
Using the fact that $(1/D)\sum_{m}(-1)^{\boldsymbol{m}\cdot(\boldsymbol{n} + \boldsymbol{k})} = \delta_{n,k}$, we have found that
\begin{align}
{a_{mn}(t) = \frac{1}{D}\sum_{k=0}^{D-1}(-1)^{\boldsymbol{\pi}(k)\cdot\boldsymbol{n} + \boldsymbol{k}\cdot\boldsymbol{m}}}
\end{align}
This is equivalent to the statement that we wish to prove.  We observe that if $|m\rangle$ is a product state in the Pauli $Z$ basis, where the spin configuration corresponds to the binary representation of the integer $m$ as in Eq. \ref{eq:1_binary_state}, then applying a Hadamard gate to every spin yields 
\begin{align}
H^{\otimes N}\ket{m} = \frac{1}{\sqrt{D}}\sum_{n=0}^{D-1}(-1)^{\boldsymbol{m}\cdot\boldsymbol{n}}|n\rangle
\end{align}
Now, we observe that 
\begin{align}
\langle n | H^{\otimes N}U(t) \,H^{\otimes N}| m\rangle &= \frac{1}{D}\sum_{k,\ell = 0}^{D-1}(-1)^{\boldsymbol{m}\cdot\boldsymbol{k} + \boldsymbol{n}\cdot\boldsymbol{\ell}}\delta_{\pi(k),\ell}\nonumber\\
&= a_{mn}(t)\nonumber
\end{align}
We conclude that 
\begin{align}
W(t) = H^{\otimes N}U(t)\,H^{\otimes N}
\end{align}

\section{Operator Entanglement Entropy}\label{appendix:op_entanglement}
We calculate the purity of the wavefunction in Eq. (\ref{eq:n}), which quantifies the complexity of the Pauli $Z$ operator spreading from the automaton dynamics.  Consider a bi-partitioning of the state $|n(t)\rangle$ in to an $A$ and $B$ subsystem, with Hilbert space dimensions $D_{A}$ and $D_{B} = D/D_{A}$, respectively.  Furthermore, let $\{|m_{A, B}\rangle\}$ denote a complete set of states in $A$ and $B$, respectively; as before, we let $\boldsymbol{m}_{A, B}$ denote the binary vector representation of these states, so that the dot product $\boldsymbol{n}\cdot\boldsymbol{m} = \boldsymbol{n}_{A}\cdot\boldsymbol{m}_{A} + \boldsymbol{n}_{B}\cdot\boldsymbol{m}_{B}$.  

Using this notation, we may write down the reduced density matrix as 
\begin{align}
\rho_{A}(n) &= \Tr_{B} |n(t)\rangle\langle n(t)|\\
&= \sum_{m_{A},\,k_{A} = 0}^{D_{A}-1} \rho_{A}(m_{A}, k_{A}) \,|m_{A}\rangle\langle k_{A}|,\nonumber
\end{align} 
where the matrix elements of the reduced density matrix are given by
\begin{align}
&\rho_{A}(m_{A}, k_{A}) = \sum_{m_{B}, k_{B} = 0}^{D_{B}-1} \delta_{m_{B}, k_{B}}\,a_{nm}(t) a_{nk}(t).
\end{align}
Substituting the expression for $a_{nm}(t)$ and performed the sum over $m_{B}$, $k_{B}$ in the above expression yields
\begin{align}
\rho_{A}(m_{A}, k_{A}) = \frac{D_{B}}{D^{2}}\sum_{q, \,r = 0}^{D-1}&\Big[\delta_{q_{B}, r_{B}}(-1)^{\boldsymbol{q}_{A}\cdot\boldsymbol{m}_{A} + \boldsymbol{r}_{A}\cdot\boldsymbol{k}_{A}}\\
&\times (-1)^{\boldsymbol{n}\cdot\left[\boldsymbol{\pi^{-1}}(q) + \boldsymbol{\pi^{-1}}(r)\right]}\Big].\nonumber 
\end{align}
Here, $\pi^{-1}$ is the inverse of the permutation $\pi\in S_{D}$, while $\boldsymbol{\pi^{-1}}(q)$ is the vector representation of the element $\pi^{-1}(q)$, as before.  Using this expression, we find that the purity (exponential of the second R\'{e}nyi entropy is given by
\begin{align}\label{eq:purity_appendix}
\Tr\,\rho_{A}^{\,2}\, &= \frac{1}{D^{2}}\sum_{r,\,r',\,q,\,q' = 0}^{D-1}\Big[ \delta_{r_{A}, q_{A}}\delta_{r'_{A}, q'_{A}}\delta_{r_{B}, r'_{B}}\delta_{q_{B},q'_{B}}\nonumber\\
&\times (-1)^{\boldsymbol{n}\cdot\left[\boldsymbol{\pi^{-1}}(q) + \boldsymbol{\pi^{-1}}(q') + \boldsymbol{\pi^{-1}}(r) + \boldsymbol{\pi^{-1}}(r') \right]}\Big].
\end{align}

We now calculate the purity, averaged over all permutations $\pi\in S_{D}$.  In this case, we observe that for non-zero $\boldsymbol{n}$, the quantity $\boldsymbol{n}\cdot\boldsymbol{\pi^{-1}}(q)$ can be an even or odd integer with equal probability, for a randomly chosen permutation $\pi$.  Therefore, for non-zero $\boldsymbol{n}$, we observe that 
\begin{align}
\frac{1}{D!}\sum_{\pi\in S_{D}}(-1)^{\boldsymbol{n}\cdot[\boldsymbol{\pi^{-1}}(q) + \boldsymbol{\pi^{-1}}(q')]} = \delta_{qq'}.
\end{align}
Similarly, the average of the expression in Eq. (\ref{eq:purity_appendix}) only contributes if pairs of elements are equal.  Therefore for non-zero $\boldsymbol{n}$, 
\begin{align}\label{eq:permutation_avg}
&\frac{1}{D!}\sum_{\pi\in S_{D}}(-1)^{\boldsymbol{n}\cdot[\boldsymbol{\pi^{-1}}(q) + \boldsymbol{\pi^{-1}}(q') + \boldsymbol{\pi^{-1}}(r) + \boldsymbol{\pi^{-1}}(r')]} \nonumber\\
&= \delta_{qq'}\delta_{rr'} + \delta_{qr}\delta_{q'r'} + \delta_{qr'}\delta_{q'r} - 2\,\delta_{qq'}\delta_{rr'}\delta_{qr}.
\end{align}

We may use this expression to compute the purity \emph{averaged} over all choices of permutations, which we denote as $\overline{\,\Tr[\rho_{A}^{2}]\,}$.  Substituting Eq. (\ref{eq:permutation_avg}) into Eq. (\ref{eq:purity_appendix}) yields the result that 
\begin{align}
\overline{\,\Tr\,\rho^{2}_{A}(n)\,} = \left\{ \begin{array}{cc} 1 & (n=0)\\
&\\
\displaystyle{D_{A}^{-1} + D_{B}^{-1} - D^{-1}} & (n\ne 0)
\end{array}\right..
\end{align}

\section{Recurrence Time Distribution}\label{appendix:permutation_group}
Consider the permutation group $S_{D}$.  If $\pi\in S_{D}$ has a length-$\ell$ cycle that includes a particular element (say the element $1$), then we may write $\pi$ as the product of two commuting permutations, i.e. 
\begin{align}
\pi = (1\,k_{2}\,\ldots\,k_{\ell}) \,\sigma,
\end{align}
with $\sigma \in S_{D-\ell}$ acting on the $D-\ell$ elements that do not appear in the length-$\ell$ cycle.  For a fixed set of elements $k_{2},\ldots,k_{\ell}$, the number of such permutations $\pi\in S_{D}$ is precisely $(\ell-1)! (D-\ell)!$.  Furthermore, the number of ways to choose these elements out of $D$ elements is $\left(\begin{array}{c}D-1\\ \ell-1\end{array}\right)$.  Therefore, the number of permutations $\pi\in S_{D}$ with a length-$\ell$ cycle that contains the element $1$ is precisely
\begin{align}
(\ell-1)! (D-\ell)! \left(\begin{array}{c}D-1\\ \ell-1\end{array}\right) = (D-1)!\,.
\end{align}
Alternatively, the probability that a random permutation in $S_{D}$ has a length-$\ell$ cycle containing a particular, fixed element $m$ is 
\begin{align}
P_{\ell, m} = \frac{1}{D}.
\end{align}  

For a permutation $\pi\in S_{D}$, we define $\ell_{n}(\pi)$ to be the length of the cycle that element $n$ is in.  Using this, we may write that the average number of cycles $\langle{N}\rangle$ in a permutation is
\begin{align}
\langle{N}\rangle = \frac{1}{D!}\sum_{\pi\in S_{D}} \left[\sum_{n=1}^{D}\frac{1}{\ell_{n}(\pi)}\right] = \sum_{n=1}^{D}\left[\sum_{\ell=1}^{D} \frac{P_{\ell, n}}{\ell}\right] = \sum_{\ell=1}^{D}\frac{1}{\ell}.\nonumber
\end{align}
Similarly, the average return time for a random element $\langle t \rangle$ is given by
\begin{align}
\langle t \rangle &= \frac{1}{D!}\sum_{\pi\in S_{D}} \left[\frac{1}{D}\sum_{n=1}^{D}{\ell_{n}(\pi)}\right]\nonumber\\
&= \frac{1}{D}\sum_{n=1}^{D}\left[\sum_{\ell=1}^{D} P_{\ell,n}\,\ell\right] = \frac{D+1}{2}.
\end{align}
Therefore, we conclude that 
\begin{align}
\langle t \rangle\overset{D\rightarrow\infty}{=} \frac{D}{2} \hspace{.4in} \langle{N}\rangle \overset{D\rightarrow\infty}{=}  \log(D).
\end{align}

\section{Dynamics with 1D Subsystem Symmetries in Two Dimensions}\label{appendix:1d_subsys_dynamics}
We assume that the correlation function $G_{\rB, t}\equiv \Tr\left[Z_{\rB}(t)Z_{0}(t)\right]/D$, 
where $D = 2^{N}$ is the Hilbert space dimension of the $N$-site system evolves according to the difference equation, 
\begin{align}
&G_{\rB,t+1} = \frac{3}{4}G_{\rB,t} + \frac{1}{8}\left[G_{\rB + \hat{x},t}+G_{\rB - \hat{x},t}+G_{\rB + \hat{y},t}+G_{\rB - \hat{y},t}\right] \nonumber\\
&- \frac{1}{16}\Big[G_{\rB + \hat{x}+\hat{y},t}+G_{\rB - \hat{x}-\hat{y},t} +G_{\rB -\hat{x} + \hat{y},t}+G_{\rB +\hat{x}- \hat{y},t}\Big].\nonumber
\end{align}
Taking the continuum limit in time, so that $G_{\rB, t+1} \approx G(\rB, t) + dG(\rB ,t)/dt$, and going to momentum space, 
\begin{align}
G(\kB, t) \equiv \frac{1}{\sqrt{N}}\sum_{\rB}e^{i\kB\cdot\rB}G(\rB, t)
\end{align} 
we find that
\begin{align}
\frac{d}{dt}G(\kB, t) = -f(\kB)G(\kB, t),
\end{align}
where
\begin{align}
f(\kB) \equiv \sin^{2}\left(\frac{k_{x}}{2}\right)\sin^{2}\left(\frac{k_{y}}{2}\right),
\end{align}
which then gives the solution
\begin{align}
G(\rB, t) = \frac{1}{N}\sum_{\kB}e^{-i\kB\cdot\rB - f(\kB)t}
\end{align}
given the initial condition $G(\rB, 0) = \delta_{\rB,0}$. 

We now evaluate $G(0,t)$: in the thermodynamic limit, we make the replacement $N^{-1}\sum_{\kB} \rightarrow \int d^{2}\kB/(2\pi)^{2}$, so that
\begin{align}
G(0,t) = \int\frac{d^{2}\kB}{(2\pi)^{2}} e^{ - f(\kB)t}.
\end{align}
Performing the integral over $k_{x}$ yields
\begin{align}
G(0,t) &= \int_{\text{-}\pi}^{\pi}\frac{dk_{y}}{2\pi}\exp\left[-\frac{t}{2}\sin^{2}\left(\frac{k_{y}}{2}\right)\right]\,I_{0}\left[\frac{t}{2}\sin^{2}\left(\frac{k_{y}}{2}\right)\right]
\end{align}
where $I_{0}(z)$ is the modified Bessel function of the first kind.  Finally, the integral over $k_{y}$ yields 
\begin{align}
{G(0,t) = \,_{\frac{1}{2}}F_{\frac{1}{2}}(1; 1; -t)},
\end{align}
where $\,_{a}F_{b}(p;q;z)$ is the generalized hypergeometric function.   The asymptotic expansion of $G(0,t)$ at long times is then
\begin{align}
{G(0,t) \overset{t\rightarrow\infty}{=} \frac{\ln(t)}{\pi^{3/2}\sqrt{t}} + O\left(\frac{1}{\sqrt{t}}\right)}.
\end{align}
More precisely, the asymptotic expansion is
\begin{align}
G(0,t)= \frac{1}{\pi^{3/2}\sqrt{t}}\left[\ln(t) - 2\gamma - \frac{3}{\sqrt{\pi}}\Gamma'(1/2)\right] + \cdots \, .
\end{align}
Resolving the $\ln(t)/\sqrt{t}$ behavior would require going to very long times (note the logarithm becomes approximately twice as large as the constant correction only when $t \sim 10^{4}$).

We now determine the late-time scaling of the correlation function $G(\rB, t)$
\begin{align}
G(\rB,t) = \int\frac{d^{2}\kB}{(2\pi)^{2}} e^{ i\kB\cdot\rB - f(\kB)t}
\end{align}
along the line $\rB = (x, 0)$.  Hereafter, we refer to this correlation function as $G(x, t)$. Performing the integral over $k_{y}$, we obtain
\begin{align}\label{eq:G}
G(x,t) &= 2\int_{0}^{\pi}\frac{dk}{2\pi}\cos(kx)\, g(k,t)
\end{align},
where 
\begin{align}
g(k,t) \equiv e^{-\frac{t}{2}\sin^{2}\left(\frac{k}{2}\right)}\,I_{0}\left[\frac{t}{2}\sin^{2}\left(\frac{k}{2}\right)\right].
\end{align}
At long times $t \gg 1$, the quantity  $(t/2)\sin^{2}(k/2)$ is large over the interval $k \gtrsim 1/\sqrt{t}$.  The scaling of the correlation function in this regime may be approximated by replacing $g(k,t)$ by its asymptotic form at long times, for small $k$,
\begin{align}
\int_{0}^{\pi}\frac{dk}{2\pi}\cos(kx)g(k,t) &\overset{t\gg 1}{\sim} \frac{1}{\sqrt{t}}\int_{k\gtrsim \sqrt{t}}\frac{dk}{2\pi}\frac{ \cos(kx)}{k}. \nonumber
\end{align}
This integral is logarithmically divergent at small $k$.  When $t \gg x^{2}\gg 1$, we may evaluate the integral perturbatively in $x/\sqrt{t}$ to find that
\begin{align}
G(x,t)\sim \frac{1}{\sqrt{t}}\left[\log\left(\frac{\sqrt{t}}{x}\right) + O(1)\right].
\end{align}
So that the correlation function ``peaks" when $t\sim x^{2}$, and the maximum peak height scales as $\sim 1/x$. 

\section{Line-Like Symmetries in 3D}\label{appendix:line_symm_3d}
For a three-dimensional system with intersecting, line-like symmetries, we consider the natural generalization of Eq. (\ref{eq:double_laplacian}) in the space-time continuum limit:
\begin{align}
\partial_{t}G(\rB, t) = \lambda\,\partial_{x}^{2}\partial_{y}^{2}\partial_{z}^{2}G(\rB, t)
\end{align}
where $\lambda$ is a free parameter of the evolution, so that 
\begin{align}
G(\rB, t) &\sim \int d^{3}\kB\,\exp\left[{i\kB\cdot\rB - k_{x}^{2}k_{y}^{2}k_{z}^{2}t}\right]\\
&= \frac{2\pi^{3/2}}{\sqrt{t}}\,\mathcal{G}^{5,0}_{2,7}\left(\begin{array}{c} \boldsymbol{v}\\
\boldsymbol{w}
\end{array} \Bigg| \frac{x^{2}y^{2}z^{2}}{64\lambda t}\right),
\end{align}
where $\mathcal{G}^{m,n}_{p,q}$ is the Meijer G-function, and the vectors $\boldsymbol{v}$ and $\boldsymbol{w}$ are defined as 
\begin{align}
\boldsymbol{v} \equiv \left( \frac{1}{4},\, \frac{3}{4}\right)\hspace{.2in}
\boldsymbol{w} \equiv \left(0,\,0,\,0,\, \frac{1}{2},\, \frac{1}{2},\, \frac{1}{4},\, \frac{3}{4}\right).
\end{align}
The asymptotic form $t \gg x^{2}y^{2}z^{2}$  of this integral yields the leading, long-time behavior
\begin{align}
G(\rB, t) {\sim} \frac{\log^{2}(t) + O(\log(t))}{\sqrt{t}} \hspace{.25in}(\lambda t \gg x^{2}y^{2}z^{2}).
\end{align}

\bibliography{library}

\end{document}